\documentclass[twocolumn,notitlepage,prl,superscriptaddress,longbibliography]{revtex4-2}
\usepackage{bbm}
\usepackage{amssymb}
\usepackage{algpseudocode}
\usepackage{algorithm}
\usepackage{amsmath}
\usepackage{braket}
\usepackage{booktabs}
\usepackage{color}
\usepackage{comment}
\usepackage{dsfont}
\usepackage{etoolbox}
\usepackage{epstopdf}
\usepackage[T1]{fontenc}
\usepackage{float}
\usepackage{graphicx}
\usepackage[colorlinks=true,linkcolor=blue,citecolor=blue,urlcolor=blue]{hyperref}
\usepackage[latin9]{inputenc}
\usepackage{indentfirst}
\usepackage{latexsym,bm,euscript}
\usepackage{mathrsfs}
\usepackage{multirow}
\usepackage{natbib}
\usepackage{soul}
\usepackage{subfigure}
\usepackage{times}
\usepackage{titlesec}
\usepackage{tikz}
\usepackage{txfonts}
\usepackage{xspace}
\usepackage{xfrac}
\usepackage{etoolbox}
\usepackage{tikzMPS} 
\usepackage{tikz-cd}
\usepackage{physics}
\usepackage{xfp}
\usepackage[normalem]{ulem}
\usepgflibrary {shadings}
\usetikzlibrary {arrows.meta}

\renewcommand{\Tr}[1]{\mathrm{Tr}\left\{#1\right\}}
\newcommand{\kett}[1]{\left|\left.#1 \right \rangle \right \rangle}
\newcommand{\bbra}[1]{\left\langle \left\langle#1 \right.\right|}

\definecolor{DarkBlue}{RGB}{50,120,180}
\definecolor{TextBlue}{RGB}{30,70,150}

\begin{document}
\title{Tangent Space Approach for Thermal Tensor Network Simulations of the 2D Hubbard Model}

\author{Qiaoyi Li}
\affiliation{School of Physics, Beihang University, Beijing 100191, China}
\affiliation{CAS Key Laboratory of Theoretical Physics, 
Institute of Theoretical Physics, Chinese Academy of Sciences, Beijing 100190, China}
\affiliation{Hefei National Laboratory, University of Science and Technology of China, Hefei 230088, China}

\author{Yuan Gao}
\affiliation{School of Physics, Beihang University, Beijing 100191, China}
\affiliation{CAS Key Laboratory of Theoretical Physics, 
Institute of Theoretical Physics, Chinese Academy of Sciences, Beijing 100190, China}

\author{Yuan-Yao He}
\affiliation{Institute of Modern Physics, Northwest University,
Xi'an 710127,China 
}
\affiliation{Shaanxi Key Laboratory for Theoretical Physics Frontiers,
Xi'an 710127,China 
}
\affiliation{Hefei National Laboratory, University of Science and Technology of China, Hefei 230088, China}

\author{Yang Qi}
\email{qiyang@fudan.edu.cn}
\affiliation{State Key Laboratory of Surface Physics and Department of Physics, 
Fudan University, Shanghai 200433, China}
\affiliation{Collaborative Innovation Center of Advanced Microstructures, 
Nanjing 210093, China}
\affiliation{Hefei National Laboratory, University of Science and Technology of China, Hefei 230088, China}

\author{Bin-Bin Chen}
\email{bchenhku@hku.hk}
\affiliation{Department of Physics and HKU-UCAS Joint Institute of 
Theoretical and Computational Physics, The University of Hong Kong,
Pokfulam Road, Hong Kong, China}

\author{Wei Li}
\email{w.li@itp.ac.cn}
\affiliation{CAS Key Laboratory of Theoretical Physics, Institute of 
Theoretical Physics, Chinese Academy of Sciences, Beijing 100190, China}
\affiliation{Hefei National Laboratory, University of Science and Technology of China, Hefei 230088, China}
\affiliation{Peng Huanwu Collaborative Center for Research and Education, 
Beihang University, Beijing 100191, China}
\affiliation{CAS Center for Excellence in Topological Quantum Computation, 
University of Chinese Academy of Sciences, Beijng 100190, China}

\begin{abstract} 
    Accurate simulations of the two-dimensional (2D) Hubbard model constitute one 
    of the most challenging problems in condensed matter and quantum physics. 
    Here we develop a tangent space tensor renormalization group (tanTRG) 
    approach for the calculations of the 2D Hubbard model at finite temperature. An 
    optimal evolution of the density operator is achieved in tanTRG with a mild $O(D^3)$ 
    complexity, where the bond dimension $D$ controls the accuracy. With the tanTRG approach
    we boost the low-temperature calculations of large-scale 2D Hubbard systems 
    on up to a width-8 cylinder and $10\times10$ square lattice. For the half-filled Hubbard 
    model, the obtained results are in excellent agreement with those of determinant 
    quantum Monte Carlo (DQMC). Moreover, tanTRG can be used to explore the 
    low-temperature, finite-doping regime inaccessible for DQMC. The calculated charge 
    compressibility and Matsubara Green's function are found to reflect the strange metal 
    and pseudogap behaviors, respectively. The superconductive pairing susceptibility is 
    computed down to a low temperature of approximately $1/24$ of the hopping energy, 
    where we find $d$-wave pairing responses are most significant near the optimal doping. 
    Equipped with the tangent-space technique, tanTRG constitutes a well-controlled, 
    highly efficient and accurate tensor network method for strongly correlated 2D lattice 
    models at finite temperature.
\end{abstract}

\date{\today}

\maketitle

\textit{Introduction.---} 
The paradigmatic Hubbard model~\cite{Hubbard1963a,Gutzwiller1963a} 
is arguably the most intensively studied lattice model for strongly 
correlated electrons~\cite{Arovas2022Review,Qin2022Review}. 
It has been widely believed to capture the quintessence of 
high-temperature superconductivity~\cite{Cuprate1986,
TsueiRMP2000,Lee2006,Keimer2015Nature,Proust2018Review}, 
and recently also realized in optical lattice quantum simulations
\cite{Parsons2016,Greif2016,Cheuk2016b,Mazurenko2017,Koepsell2021,
Bloch2022}. The intriguing interplay between the spin and charge 
degrees of freedom in the Hubbard model may give rise to abundant,
even a plethora of electron orders in the finite-temperature phase diagram~\cite{Huang2019SM,
Wietek2021PRX,Georges2009FARC,WuPRX2018,Scheurer2018PNAS,Xiao2023PRX}. However, large-scale simulations of the 2D Hubbard 
model with a broad range of doping and down to low temperature yet
constitute a widely open and truly challenging problem~\cite{Qin2022Review}. 

Tensor networks (TNs) and their renormalization group methods provide powerful 
approaches for quantum many-body problems~\cite{White1992DMRG,
Verstraete2004,Schollwock2011MPS,Cirac2021RMP}. In particular, 
thermal TNs~\cite{Feiguin2005,White2009METTS,Stoudenmire2010,
Chen2018,Li2011,Czarnik2012PEPS} have been conceived and extensively used in the studies of low-dimensional quantum magnets
\cite{Xiang1998Thermodynamics,Chen2019,Li2020TMGO,
Li2021NC,YuCPL2021,Larrea2021Nature} and recently also in 
correlated fermions at finite temperature~\cite{Czarnik2016TNR,
Chen2021SLU,Wietek2021PRX,Wietek2021PRX-II,Chen2022tbg,
Dziarmaga2022PRB}. However, the accessible system size and 
lowest temperature that fermion thermal TN methods can 
handle are still rather limited. For a comparison, while 
the $T=0$ density matrix renormalization group (DMRG) 
can deal with fermion cylinders of width $W=6$-$8$
\cite{LeBlanc2015,Zheng2017,Qin2020PRX,Jiang2021PNAS}, 
finite-temperature calculations can currently reach a $W=4$ 
Hubbard cylinder~\cite{Wietek2021PRX,Wietek2021PRX-II}. 
For cracking electron secrets in the phase diagram of the 2D Hubbard model, like the strange metallicity~\cite{Huang2019SM}, 
pseudogap~\cite{Wietek2021PRX}, 
and $d$-wave superconductivity~\cite{HCJiang2019Science,
HCJiang2020PRR,Jiang2021PNAS,HCJiang2021tJ,Gong2021tJ}, 
further developments in the algorithm are highly required.

In this work, we propose a tangent space tensor renormalization group 
(tanTRG) approach for highly controlled simulations both at half filling 
and finite doping. It has the following promising features: 
(i) A versatile 2D finite-temperature approach with efficient temperature grid design.
Through a quasi-1D mapping, it systematically deals with the long-range 
interactions and evolves 2D systems based on the matrix product operator 
(MPO) representation of the Hamiltonian, making it advantageous over the
Trotter-based approach~\cite{Zwolak2004,Feiguin2005,
White2009METTS,Li2011}. 
In tanTRG we integrate 
a flow equation and have a very high degree of flexibility in designing 
temperature grids. A remarkably larger imaginary-time step can be 
taken in tanTRG compared to Trotter-based approaches.
(ii) Moderate computational complexity.
Compared to the exponential tensor renormalization group (XTRG) 
that can simulate a 2D system down to low temperatures with a relatively
high cost of $O(D^4)$~\cite{Chen2018,Chen2021SLU}, tanTRG is 
with only $O(D^3)$ complexity that allows for a significantly larger 
bond dimension $D$ in the calculations. These advantages therefore lead to 
(iii) unprecedented finite-temperature simulations of 
large-scale systems. As fermion symmetries can be conveniently 
implemented in tanTRG with the tensor library QSpace
\cite{Weichselbaum2012,Weichselbaum2020,Bruognolo2021}, 
it further reduces the computational costs and allows for up to 
$D^*=4,096$ SU(2)$_{\rm{charge}}$ $\times$ SU(2)$_{\rm{spin}}$ 
multiplets at half filling (i.e., approximately $D\approx 25,000$ 
equivalent U(1)$_{\rm charge}
\times $U(1)$_{\rm spin}$ states). For doped cases, U(1)$_{\rm charge}
\times $SU(2)$_{\rm spin}$ can also be implemented. Note the spin 
and charge symmetries can be implemented in the MPO representation
of the grand canonical ensemble (GCE) density operator. It helps further
enhance the effective bond dimension $D$ and renders excellent 
accuracy for large-scale Hubbard systems on up to a width-8 cylinder 
and $10\times10$ square lattice down to sufficiently low temperature.

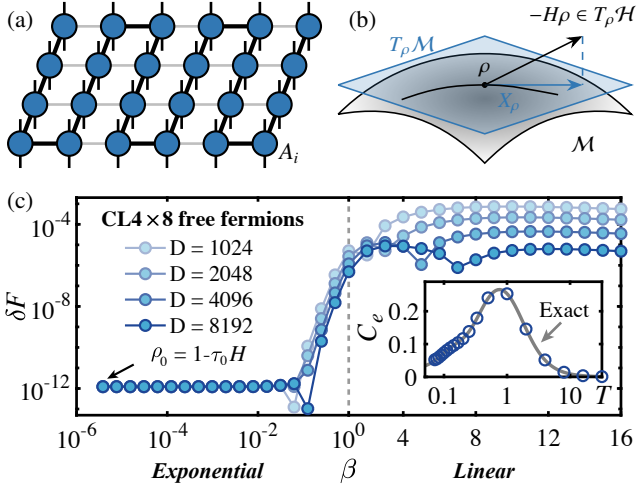
\begin{figure}
\centering
\begin{tikzpicture}[scale = 0.65]
\node at (0, 2.5) {\small(a)};
\foreach \i in {0, 1, 2, 3, 4, 5}{
\foreach \j in {3, 2, 1, 0}{
\draw [line width = 1, black]  (\i + 0.3*\j, 0.8*\j-0.5)  -- (\i + 0.3*\j, 0.8*\j + 0.5);
\ifnum \j>0 
\draw [line width = 1.6, black] (\i + 0.3*\j, 0.8*\j) -- (\i + 0.3*\j - 0.3, 0.8*\j - 0.8);
\fi
\ifnum \i < 5
\ifthenelse{
\(\j = 0 \and \(\i = 0 \OR \i = 2 \OR \i = 4\)\) \OR
\(\j = 3 \and \(\i = 1 \OR \i = 3 \OR \i = 5\)\)
}{
\draw [line width = 1.6, black] (\i + 0.3*\j, 0.8*\j) -- (\i + 0.3*\j +1, 0.8*\j);
}{
\draw [line width = 1, lightgray] (\i + 0.3*\j, 0.8*\j) -- (\i + 0.3*\j +1, 0.8*\j);
}
\fi
\draw [line width = 0.8, black, fill = DarkBlue] (\i + 0.3*\j, \fpeval{0.8*\j}) circle [radius = 0.25];
}
}

\node at (5.5, -0.2) {\small$A_i$}; 
        
\node at (7, 2.5) {\small(b)};
        \draw [line width = 0.6, inner color = gray, outer color = white] (6.5, 0.6) to 
        [out = 10, in = 135] (9.5, -0.4)
        to [out = 45, in = 170] (12.5, 0.6) to [out = 135, in = 45] (6.5, 0.6);

        \draw [line width = 0.6, DarkBlue, fill = DarkBlue, fill opacity = 0.2] 
        (6.5,1.2) -- (9.5,0.2) -- (12.5, 1.2) -- (9.5, 2.2) -- (6.5, 1.2);

        \draw [fill] (9.5, 1.2) circle [radius = 0.05];
        \draw [line width = 0.6, -{Stealth[length=6pt]}] (9.5, 1.2) -- (11.5, 2.2);
        \draw [line width = 0.6, DarkBlue, -{Stealth[length=6pt]}] (9.5, 1.2) -- (11.5, 1.2);
        \draw [line width = 0.6, dashed, DarkBlue] (11.5, 2.2) -- (11.5, 1.2);

        \draw [line width = 0.6] (7.8, 0.9) to [out = 20, in = 180] (9.5, 1.2);
        \draw [line width = 0.6] (9.5, 1.2) to [out = 0, in = 170] (11, 1);

        \node [above] at (9.5, 1.2) {\footnotesize$\rho$};
        \node at (11.5, 0) {\footnotesize$\mathcal{M}$};
        \node [above] at (8, 1.6) {\color{DarkBlue}{\footnotesize$T_{\rho}\mathcal{M}$}};
        \node [above] at (11.5, 2.2) {\footnotesize$-H\rho\in T_{\rho}\mathcal{H}$};
        \node [below, DarkBlue] at (10, 1.2) {\footnotesize$X_\rho$};
        
        \node at (0, -1.2) {\small(c)};
        \node at (6.1, -4) {\includegraphics[width = 0.97\linewidth]{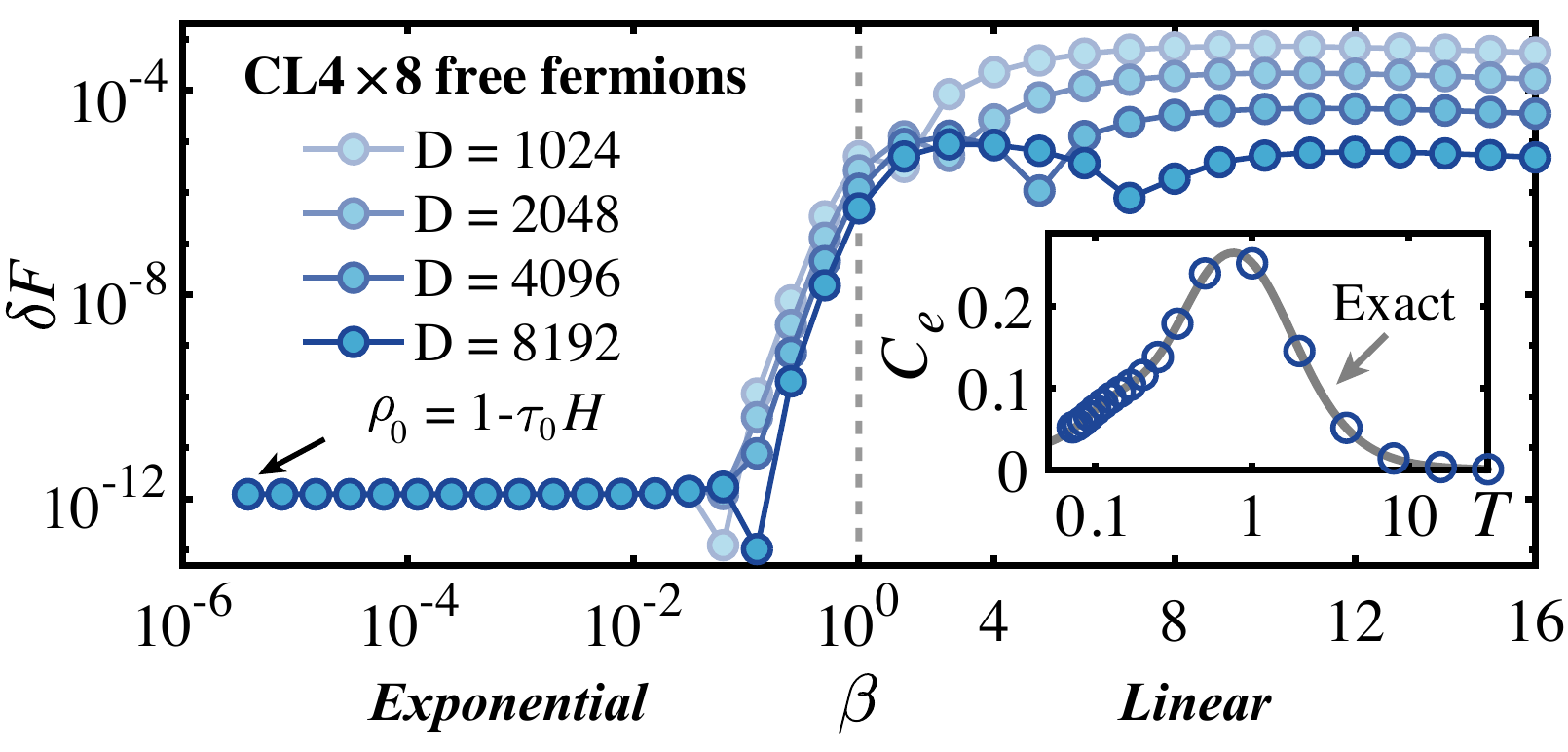}};

\end{tikzpicture}

\caption{
(a) The MPO representation of thermal density operator $\rho$ and 
corresponding quasi-1D mapping of the square lattice. The MPO consists 
of rank-4 tensors $A_i$ with two geometric and two physical indices. 
(b) The MPO manifold $\mathcal{M}$ and its tangent space $T_{\rho} 
\mathcal{M}$, where the black arrow denotes the tangent vector $-H \rho$ 
for imaginary-time evolution, and the blue one is its component within 
the tangent space $T_{\rho} \mathcal{M}$. The flow induced by the 
projected tangent vector field is indicated by the trajectory within the 
manifold $\mathcal{M}$. (c) The relative error $\delta F = |F 
- F_{\rm ex}|/|F_{\rm ex}|$ (with $F_{\rm ex}$ the exact solution) for half-filled free fermions on a $4\times8$ 
cylinder. A high accuracy is obtained by a hybrid cooling scheme with 
both exponential ($\beta \leqslant 1$) and linear ($\beta > 1$) temperature 
grids. There are two dips in $\delta F$ that represent cancellation points 
between different types of errors (see analysis in SM \cite{Supplementary}),
and the inset shows the specific heat $C_e$ in excellent agreement with the
exact solution. 
} 
\label{Fig1}  
\end{figure}

\textit{Tangent space tensor renormalization group.---} 
Finite-temperature properties are determined by the (unnormalized) 
density operator $\rho = e^{-\beta  H}$ as illustrated in Fig.~\ref{Fig1}(a), 
with $\beta$ the inverse temperature. The imaginary-time evolution 
equation reads ${d\rho}/{d\beta} = - H \rho$, with $-H \rho$ the tangent 
vector in $T_\rho\mathcal{H}$ (i.e., the tangent space of the full Hilbert
space $\mathcal{H}$), which, in general, sticks out of the tangent space 
$T_{\rho}\mathcal{M}$ of the MPO manifold $\mathcal{M}$ [see 
Fig.~\ref{Fig1}(b)], i.e., the MPO representation of $\rho$ 
will increase its bond dimension $D$ in the course of induced flow. 
In conventional thermal TN methods~\cite{Zwolak2004,Feiguin2005,
Li2011,Chen2018,Li2011,Czarnik2012PEPS}, the so-called truncation 
process is introduced to bring the evolved MPO back to manifold 
$\mathcal{M}$ with a fixed $D$. 

Alternatively, here we propose to optimize $\rho$ within the MPO manifold 
$\mathcal{M}$ using the technique of the time-dependent variational principle 
(TDVP)~\cite{TDVP2011,MPSManifold2014,TDVP2016,MingruYang2020,
GeometryTDVP2020}, which was originally conceived for real-time evolutions 
of pure quantum states. For a generalization to density operator $\rho$, we find the 
optimal tangent vector $X_\rho$ on the tangent space $T_{\rho}\mathcal{M}$, i.e., 
\begin{equation}
\frac{d\rho}{d\beta} = 
\mathop{\arg\min}_{X_\rho\in T_{\rho}\mathcal{M}} \left\| X_\rho + H {\rho}\right\|,
\label{Eq:TanVector}
\end{equation} 
which defines a tangent vector field $\rho \mapsto X_\rho$ that induces 
the flow of $\rho(\beta)$ exactly on the manifold $\mathcal{M}$. With the 
MPO parameterization of $\rho$, the imaginary-time flow equation 
can be expressed with local tensors (c.f., Supplemental Material (SM)~\cite{Supplementary})
\begin{equation}
\frac{dA_i}{d\beta} = - H_i^{(1)} A_i^{\,} + A_i^L H_i^{(0)} S_i^{\,},
\label{Eq:TDVP}
\end{equation} 
where $H^{(1)}_i$ is the one-site effective Hamiltonian 
acting on the on-site tensor $A_i$, and $H_i^{(0)}$ 
is the bond effective Hamiltonian acting on the bond tensor $S_i$.

Following the splitting method~\cite{GeometricIntegral}, 
we separate Eq.~(\ref{Eq:TDVP}) into two linear equations $dA_i/d\beta = 
-H_{i}^{(1)}A_i$ and $dS_i/d\beta = H_i^{(0)}S_i$ regarding the site and bond 
updates, respectively, and then integrate the equations sequentially 
in a sitewise sweep to conduct the time evolution. 
Taking a left-to-right sweep as an example, we first update 
the local tensor $A_i(\beta_0+\tau) = e^{-\tau H_i^{(1)}} A_i(\beta_0)$ with
the Lanczos-based exponential method, then left-canonicalize $A_i$ via a 
QR decomposition $A_i = A_i^L S_i$. Subsequently, we conduct backward 
evolution of bond tensor $S_i(\beta_0+\tau) = e^{\tau H_i^{(0)}} S_i(\beta_0)$,
associate it to $A_{i+1}$, and then move on to the next site. Such a sweep 
process naturally maintains the canonical form of the MPO~\cite{Supplementary}, 
and guarantees an optimal approximation within its manifold $\mathcal{M}$.

\begin{figure}
\centering
\includegraphics[width = 1\linewidth]{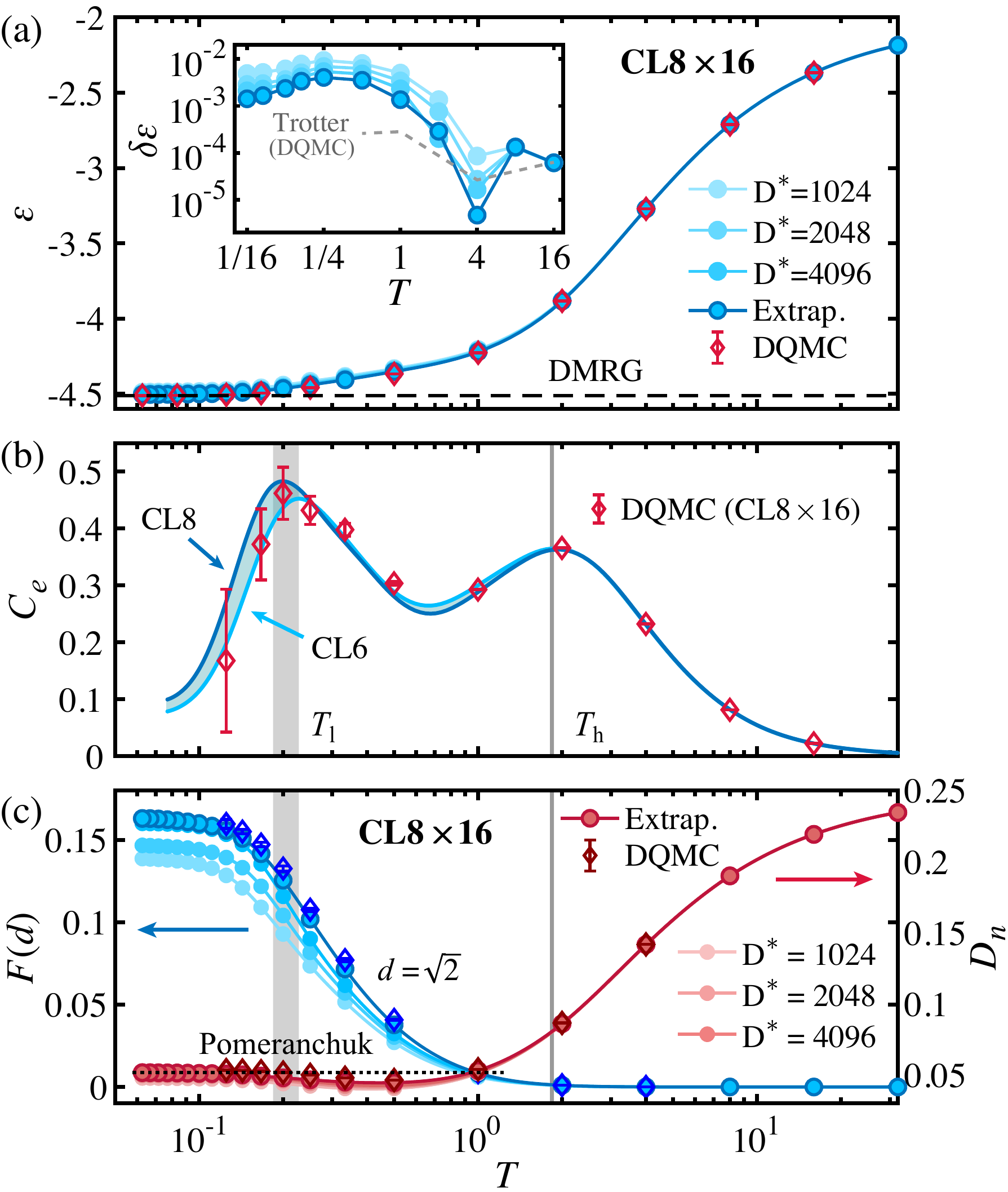}
\caption{(a) The results of the half-filled Hubbard model. The relative 
difference $\delta \varepsilon = \tfrac{|\varepsilon - \varepsilon_\mathrm{QMC}|}{|\varepsilon_\mathrm{QMC}|}$ of the tanTRG results (up to $D^*=4096$ 
and extrapolated to infinite $D$) are plotted vs $T$ in the inset, with the 
estimated Trotter errors of DQMC also indicated. The ground-state energy 
is obtained by standard two-site DMRG with $D^\ast = 8192$ (truncation error $\lesssim 10^{-5}$). 
(b) The electron specific heat $C_e$ of 
CL$6\times12$ and CL$8\times16$, where the two $C_e$ peaks 
indicate two temperature scales, namely, $T_{\rm h}$ and $T_{\rm l}$.
(c) The double occupancy $D_n$ and spin-spin correlations 
$F(d)$ with $d \equiv \sqrt{2}$, which change rapidly near $T_{\rm h}$ 
and $T_{\rm l}$, respectively. The anomalous decrease in $D_n$ near 
$T_{\rm l}$ as $T$ rises reflects the Pomeranchuk effect in the Hubbard 
model.
}
\label{Fig2}
\end{figure}

\textit{2D Hubbard model on the square lattice.---}
We consider the single-band Hubbard model on a square lattice, 
whose Hamiltonian reads
\begin{equation}
H = -t\sum_{\langle i,j\rangle,\sigma}\left(c_{i\sigma}^\dagger 
c_{j\sigma}^{\phantom{\dagger}} + {\rm H.c.}\right) + 
U\sum_{i}n_{i\uparrow}n_{i\downarrow} -\mu\sum_{i} n_i,
\label{Eq:Hubbard}
\end{equation}
where $t=1$ is chosen as the energy scale, and $\mu$ controls the fermion filling 
$n$ (or hole doping $\delta = 1 - n$). The on-site repulsion is fixed as $U=8$ 
if not otherwise mentioned. The calculations are performed on the 
cylinder lattice (CL) wrapped around the circumference direction (width $W$) 
while left open along the longitudinal direction (length $L$), and also open 
square (OS) lattice with full open boundaries.

\textit{Benchmarks on noninteracting fermions.---}
We start with benchmarks on free fermions with $U=0$. 
The tanTRG calculations can be initialized from a high-temperature 
density operator $\rho_0 = 1 - \tau_0 H$ with very small $\tau_0 \sim 
10^{-6}$, where a compact representation of $\rho_0$ can 
be conveniently constructed from the MPO representation of the Hamiltonian
\cite{Frowis2010Tensor,Pirvu2010MPO,Hubig2017MPO,SETTN,XTRG}. 
After that, we cool down the system by integrating the flow equation, 
Eq.~(\ref{Eq:TDVP}), following flexible temperature grids, and 
compute the finite-temperature properties from $\rho(\beta)$. 

In practice, we always start with exponential grids and exploit the two-site 
update allowing the MPO bond dimension $D$ to increase adaptively.
Successively, a pretty large and constant step length $4\tau = 1$ is 
adopted in the linear evolution stage. Very accurate results in free 
energy and specific heat are obtained in Fig.~\ref{Fig1}(c), as not 
only the projection but also Lie-Trotter errors are well controlled by 
bond dimension $D$~\cite{Supplementary}. Remember that the free 
fermion system, though being exactly soluble, poses challenges for 
TN methods due to the high entanglement associated with the Fermi surface 
(FS). Here the accurate results on free fermions show that tanTRG 
provides a powerful tool for tackling more realistic problems.

\textit{2D Hubbard model at half filling.---} 
In Fig.~\ref{Fig2}, we present the tanTRG results on a width-8 cylinder 
CL$8\times16$, and leave the results on narrower cylinders ($W=4,6$) 
to the SM~\cite{Supplementary}. In practical calculations, we expand 
$\rho_0$ to higher orders~\cite{SETTN} with a slightly larger $\tau_0 
\sim 10^{-4}$, and a bilayer technique is used to compute thermodynamic 
quantities~\cite{BilayerLTRG}. In Fig.~\ref{Fig2}(a) the results of the energy 
per site $\varepsilon$ are found in excellent agreement with the 
determinant quantum Monte Carlo (DQMC)~\cite{Blankenbecler1981,
Hirsch1983,Hirsch1985,Assaad2008} 
data down to low temperature $T/t \simeq 1/16$~\cite{Supplementary}. 

With the extrapolated $\varepsilon$ data, in Fig.~\ref{Fig2}(b) we show 
the computed specific heat $C_e= -\beta\partial \varepsilon/\partial 
\ln\beta$ again fully agrees with the DQMC results. 
In particular, the two peaks in $C_e$, i.e., $T_{\rm h}$ and $T_{\rm l}$, 
respectively, for charge and spin peaks~\cite{HubbardCe1997PRB,
Paiva2001PRB,HubbardDCA2013PRB}, constitute two-temperature 
scales. From the comparisons of CL6 and CL8 data, we find 
the higher charge peak $T_{\rm h}/t\sim 2$ has fully converged to the 
thermodynamic limit and the lower spin peak $T_{\rm l}/t\sim 0.2$ still 
changes slightly vs system widths. 

As shown in Fig.~\ref{Fig2}(c), the double occupancy $D_n = \frac{1}{N}
\sum_{i}\langle n_{i\uparrow}n_{i\downarrow}\rangle$ (with $N=L\times W$ 
the total site number) undergoes a rapid decrease at around $T_{\rm h}$, 
indicating the onset of Mott physics. Upon further cooling, the spin-spin 
correlation $F(d) = \frac{1}{N_d} \sum_{\langle i, j \rangle_d} \langle S_i 
\cdot S_{j} \rangle$ (i.e., averaged over $N_d$ pairs of sites separated by 
distance $|i-j|\equiv d$) rises up and becomes prominent below $T_{\rm l}$. 
Meanwhile, the double occupancy is found to exhibit a minimum at 
intermediate temperature $T_{\rm l} \lesssim T \lesssim T_{\rm h}$
\cite{HubbardDCA2013PRB,Paiva2010PRL}. 
This can be understood via the Maxwell's relation 
$(\partial D_n/\partial T)_U = - (\partial S/ \partial U)_T$, which associates 
the anomalous decrease in double occupancy as raising $T$ with the 
increase of magnetic entropy upon localization by enhancing $U$. This 
constitutes an intriguing quantum phenomenon in the Mott phase of 
Hubbard model~\cite{WernerPRL2005,CaiPRL2013} that resembles 
the renowned Pomeranchuk effect in $^3$He.

\begin{figure}
\centering
\includegraphics[width = 1\linewidth]{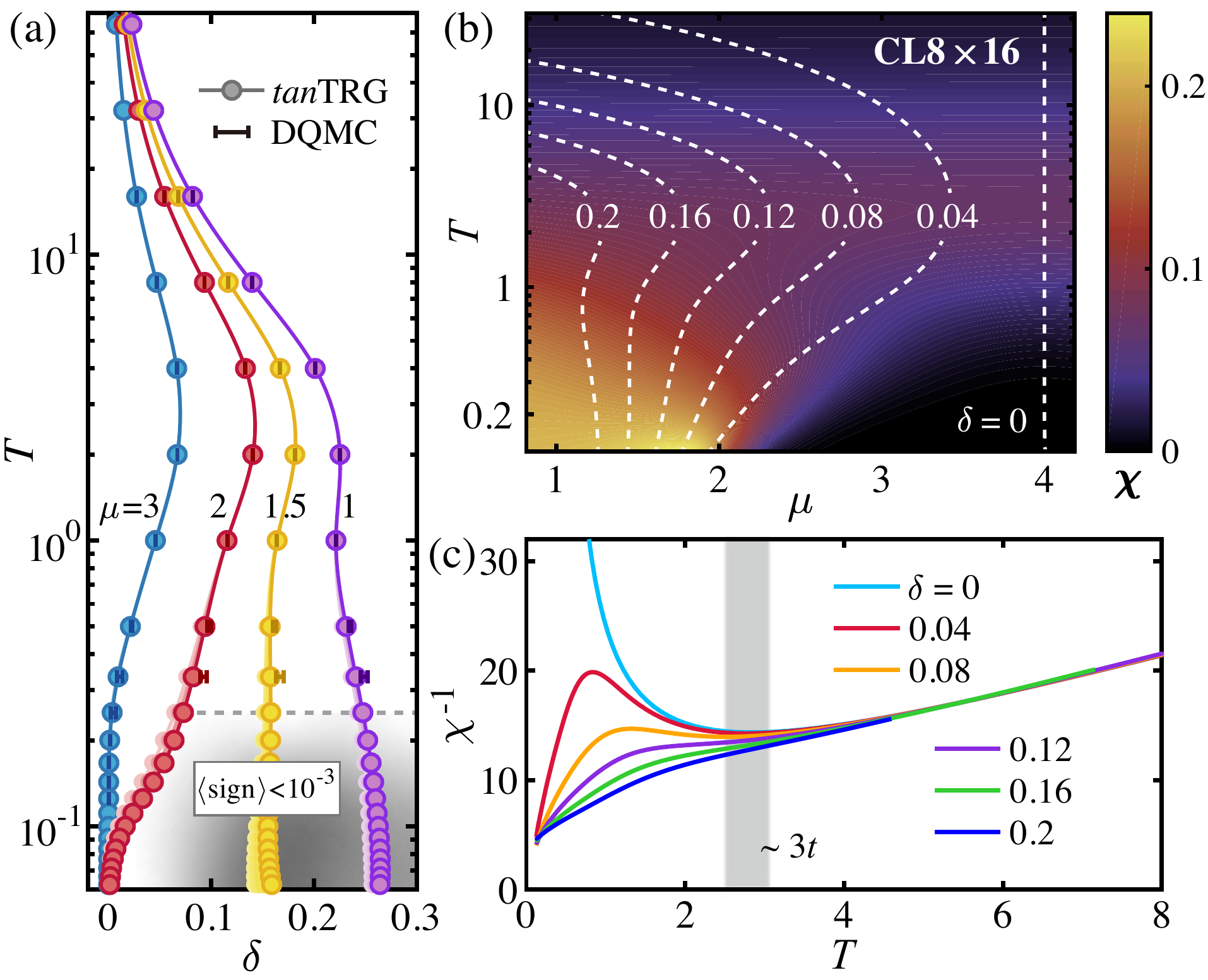}
\caption{(a) The doping $\delta$ for various $T$ and $\mu$. 
DQMC is accurate for the lightly doped case or at relatively higher 
temperature ($T/t > 0.3$) while it is hindered in the shaded regime
with $\langle \rm sign \rangle < 10^{-3}$. 
tanTRG offers accurate results even below $T/t\simeq 0.06$, 
under a wide range of dopings. The extrapolation is based on the 
$D^\ast = 1024, 2048$ and $2896$ data (and up to $D^\ast=4096$ 
for $\mu = 1.5$ case), shown as translucent symbols. 
(b) The contour plot of compressibility $\chi$, with the 
equal-$\delta$ (dashed) lines also indicated, and (c) plots the inverse 
compressibility $\chi^{-1}$ for various (interpolated) dopings $\delta$. 
}
\label{Fig3}
\end{figure}

\textit{Charge compressibility at finite doping.---} 
Now we move on to the cases with finite doping. As GCE 
is adopted in tanTRG simulations, the hole doping $\delta$ varies with 
$T$ and $\mu$ (for $\mu\neq U/2$) are shown in Fig.~\ref{Fig3}(a), 
again benchmarked with DQMC. 
For $\mu$ slightly lower than $U/2$, e.g., $\mu=3$, $\delta$ 
approaches zero in the low temperature limit and the sign problem is not 
very critical for DQMC. In contrast, when $\mu$ further deviates $U/2$ 
and the doping level increases, the DQMC sign problem becomes severe 
(i.e., $\langle \rm{sign} \rangle < 10^{-3}$, c.f., Supplemental Material
Fig.~S12~\cite{Supplementary}). 

As shown in Fig.~\ref{Fig3}(a), tanTRG produces accordant data in
the regime where DQMC works well, and can ``penetrate'' into the 
shaded low-$T$ regime inaccessible for DQMC. From Fig.~\ref{Fig3}(a) 
we note the electron density is most strongly fluctuating near $\delta 
\sim 0.1$-0.2, as evidenced by the large compressibility $\chi = 
(\partial n/\partial \mu)_T$ appearing at intermediate doping and low 
$T$ in Figs.~\ref{Fig3}(b,c). We plot the inverse compressibility $\chi^{-1}$
in Fig.~\ref{Fig3}(c) for various dopings, where the $\chi^{-1}$ results exhibit 
universal linear-$T$ behaviors for $T/t \gtrsim 3$ with little doping dependence. 
Considering that the compressibility $\chi$ has an intimate relation to 
dc resistivity via the Nernst-Einstein relation, the universal behaviors 
of $\chi^{-1}$ account for the linearity of resistivity in the high-temperature 
($T/t \gtrsim 3$) regime. Below $T/t \sim 3$, distinct $\chi^{-1}$ behaviors 
of the Mott insulator and bad metals can be clearly observed. In particular, 
$\chi^{-1}$ is found to converge to a nonzero constant for $T/t\sim 0.1$, 
which is argued to be related to the second, doping-dependent linear-$T$ 
regime of resistivity controlled instead by the diffusivity~\cite{Huang2019SM}.

\begin{figure}
    \centering
    \includegraphics[width = 1\linewidth]{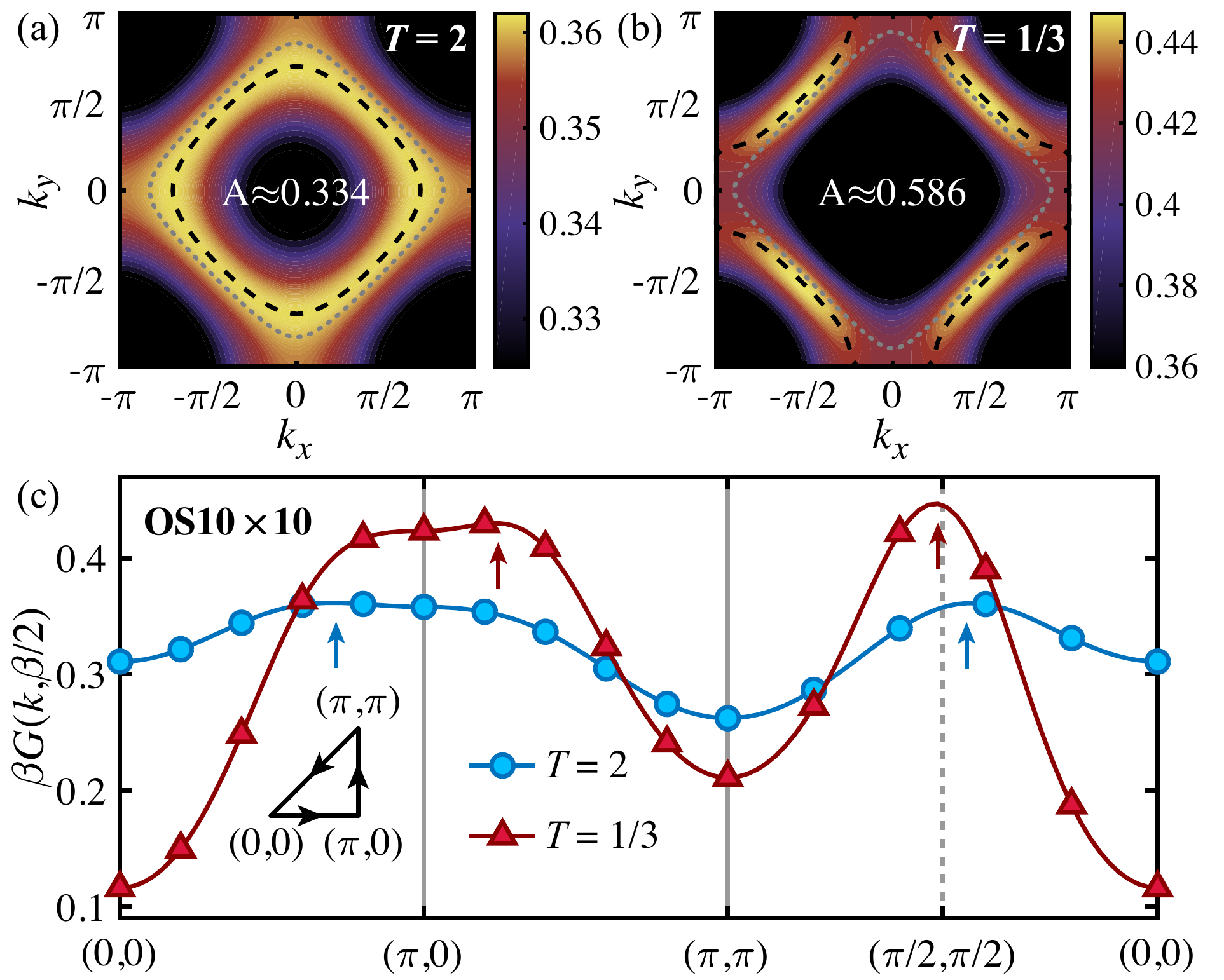}
    \caption{Topography analysis of $\beta G(k, \beta/2)$ on OS $10\times10$ 
    lattice with $\mu = 2$ for (a) $T = 2$ (with $\delta \simeq 0.14$) and (b) 
    $T = 1/3$ ($\delta \simeq 0.064$), where the high-quality data are obtained 
    with $D^\ast = 4096$. The dashed lines connect the maximum of $\beta 
    G(k, \beta/2)$ and indicate the FS, whose enclosed area is denoted as $A$ 
    [with area of the first Brillouin zone (BZ) as unit 1]. 
    The dotted line represents the FS in the noninteracting limit and with 
    the same doping. (c) The triangular path in the BZ: 
    At $T=2$ the two maxima are located in the interval $[(0,0), (\pi,0)]$ 
    and $[(\pi/2,\pi/2), (0,0)]$, indicating an electronlike FS; while for $T=1/3$, 
    the two maxima move respectively to the interval $[(\pi,0), (\pi,\pi)]$ and 
    $[(\pi,\pi), (\pi/2,\pi/2)]$ for a holelike FS. 
    }
    \label{Fig4}
\end{figure}

\begin{figure}
\centering
\includegraphics[width = 0.98\linewidth]{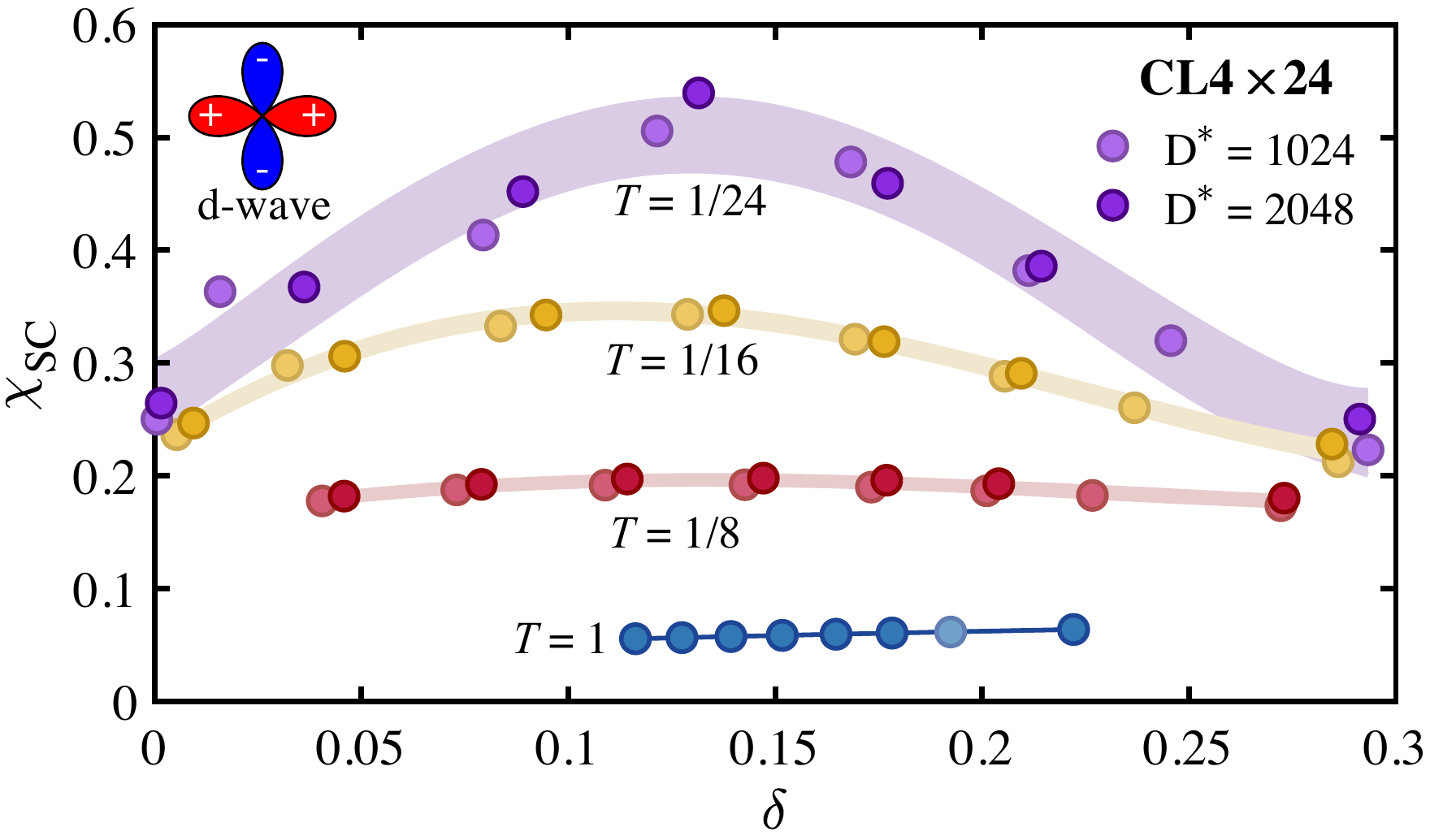}
\caption{Pairing susceptibility of the CL$4\times24$ Hubbard model with 
various dopings and temperatures. A pairing field $h_p = 0.01$ is 
adopted in the calculations, with non-Abelian Z$_{2, \rm{charge}}$ 
$\times$ SU(2)$_{\rm{spin}}$ symmetry implemented. The computed 
$\chi_{\rm SC}$ vs $\delta$ fall into the background stripes estimated 
from the polynomial fittings, whose widths represent the ($\pm 
\sigma$) confidence intervals. 
}
\label{Fig5}
\end{figure}

\textit{Matsubara Green's function and Fermi surface topology.---}  
Below the crossover temperature scale $T/t \sim 3$, the inverse compressibility 
$\chi^{-1}$ exhibits a maximum in Figs.~\ref{Fig3}(b,c) for the slightly doped 
case, e.g., $\delta=0.04,0.08$, which suggests a dramatic change in the FS
upon cooling. For this we compute the single-particle Matsubara Green's 
function $G(k,\beta/2) = \sum_{\sigma}\langle e^{\beta H/2} c_{k{\sigma}}^\dagger 
e^{-\beta H/2} \, c_{k{\sigma}} \rangle_{\beta}$ with $c_{k{\sigma}} 
= \frac{1}{\sqrt{N}} \sum_{r}e^{-ikr}c_{r{\sigma}}$
{\footnote{Note that $k$ is no longer a good quantum number under the open 
boundary condition, and thus there exist $k^\prime\neq k$ s.t. $G(k, k^\prime, \beta/2) \neq 0$. 
Nevertheless, we find that the off-diagonal components are negligible, i.e., 
about $1\%$ of total weights at $T = 2$ and $8\%$ at $T = 1/3$.}} 
that reflects the spectral weight near the FS through $\beta G(k,\beta/2) 
\sim A(k,\omega = 0)$ at low temperature~\cite{Jiang2022NC,
Samuel2017PANS}. In Fig.~\ref{Fig4}, we show the results of $\beta 
G(k,\beta/2)$ in a slightly doped case, and find quite peculiar 
temperature evolution of the FS. Despite some blurring 
due to thermal fluctuations, an electronlike FS with enclosed area 
$A < 1/2$ can be observed. As the temperature ramps down, an ``interacting 
Lifshitz transition''~\cite{Georges2009FARC,LifshitzTransition2012,
Sakai2009PRL} occurs. A holelike FS with enclosed area $A>1/2$ 
appears in Fig.~\ref{Fig4}(b), with the boundary ``pushed'' outwards 
with respect to the free-fermion FS. The unexpected holelike FS seems 
to violate the Luttinger theorem and echoes the conclusion in Refs.
\cite{Maisi2013PRL,Doiron-Leyraud2017NC,WuPRX2018,
Scheurer2018PNAS} --- the FS topology change can be associated 
with the opening of a pseudogap. Moreover, we find the signature of the pseudogap gets clearer when the system size increases, and it becomes 
very prominent when a next nearest hopping $t^\prime$ is introduced
\cite{Supplementary}.

\textit{$d$-wave pairing response.---}
Next we compute the superconductive pairing responses by applying
a pairing field $-h_p \Delta_{\rm tot} \equiv -h_p {\sum_{\langle i,j\rangle}
s_{ij}\left(\Delta_{ij} + \Delta_{ij}^\dagger\right)/2}$, where $\Delta_{ij} 
= (c_{i\downarrow}c_{j\uparrow}- c_{i\uparrow}c_{j\downarrow})/\sqrt{2}$, 
and $s_{ij} = 1$(-1) for horizontal(vertical) bonds (c.f., inset in Fig.
\ref{Fig5}). The results of pairing susceptibility $\chi_{\rm SC} = 
\tfrac{1}{N_p h_p} \langle \Delta_{\rm tot} \rangle_T$ (with $N_p$ 
the bond number) are shown in Fig.~\ref{Fig5}. At relatively high
temperature, e.g., $T/t=1$ and 1/8, $\chi_{\rm SC}$ values are 
small and insensitive to dopings. However, as the temperature further 
decreases to $T/t \lesssim 1/16$, $\chi_{\rm SC}$ displays a domelike 
shape with prominent responses near optimal doping $\delta_{\rm x} 
\approx 1/8$. Moreover, the induced superconductive order parameter 
$\tfrac{1}{N_p} \langle \Delta_{\rm tot} \rangle_T$ is found to vanish 
as $h_p \rightarrow 0$, even for the lowest accessed temperature. 
Instead, charge stripes and spin correlation modulations appear 
for $T\lesssim1/32$~\cite{Supplementary}. These results, in full 
agreement with previous studies~\cite{Qin2020PRX,Jiang2022PNAS,
Hao2022PRR,Xiao2023PRX}, show that the ground-state features 
can be well captured by tanTRG calculations down to sufficiently 
low temperature.

\textit{Summary and outlook.---}
With tangent-space techniques, we evolve the density operator 
$\rho$ optimally 
on the MPO manifold and propose a powerful 
approach for exploring 2D many-electron problems. We study 
the intriguing behaviors of charge compressibility that reflect strange 
metallicity, and unveil a holelike FS in the pseudogap regime. 
The $d$-wave pairing responses are computed down to $T/t = 1/24$, 
which are otherwise rather challenging to obtain for the 2D Hubbard model.

This approach has a wide variety of features. It can reach a low-temperature 
doped regime that is inaccessible for DQMC, and the $O(D^3)$ complexity, 
together with the implementation of non-Abelian symmetries, 
enables tanTRG to deal with wide $W=8$ cylinders at finite 
temperature. This is clearly beyond the current limit of 
$W=4$~\cite{Wietek2021PRX,Wietek2021PRX-II}, where
tanTRG obtains results in agreement with minimally 
entangled typical thermal states (METTS)\cite{Wietek2021PRX,White2009METTS,Stoudenmire2010}
(see comparisons in SM~\cite{Supplementary}). 
Overall, our results close the gap between thermal TN and 
ground-state DMRG calculations in terms of system size. 
As the cylinder width $W> 4$ is important for observing 2D correlation 
physics~\cite{Chung2020,Qin2020PRX,HCJiang2019Science,
HCJiang2020PRR,Jiang2021PNAS,HCJiang2021tJ,Gong2021tJ}, 
we believe tanTRG will play an active role in exploring the 
intriguing temperature-doping phase diagrams, and help establish 
solid connections between theories of high-$T_c$ superconductivity 
with fundamental models of correlated electrons. 

\textit{Acknowledgments}.--- The authors are indebted to Dai-Wei Qu,
Shou-Shu Gong, Lei Wang, Wei Wu, Zi Yang Meng, and Tao Shi for 
helpful discussions. This work was supported by the National Natural 
Science Foundation of China (Grants No.~12222412, 11974036, 
11834014, 12047503, 12204377 and 12174068), CAS Project for Young 
Scientists in Basic Research (Grant No.~YSBR-057), and the Innovation Program for Quantum Science and Technology (under Grant No. 2021ZD0301900).  
We thank HPC at ITP-CAS and HPC2021 at the University of Hong 
Kong for the technical support and generous allocation of CPU time.

\bibliography{./main.bib}

\clearpage
\onecolumngrid
\begin{center}
\textbf{Supplemental Material for \\ Tangent Space Approach for Thermal Tensor Network Simulations of 2D Hubbard Model}
\end{center}
\setcounter{equation}{0}
\setcounter{figure}{0}
\setcounter{table}{0}
\setcounter{page}{1}
\makeatletter
\renewcommand{\thetable}{S\arabic{table}}
\renewcommand{\theequation}{S\arabic{equation}}
\renewcommand{\thefigure}{S\arabic{figure}}
\setcounter{secnumdepth}{3}

\section{Supplementary tanTRG Results}

Below we show more benchmark results of the Hubbard model on the width $W=4,6$ 
cylinders, which turn out to be less challenging for tanTRG than the $W=8$ case 
shown in Fig.~2 of the main text. In Sec.~\ref{Subsec:Cylinder} we show the simulated 
internal energy, specific heat, and entropy results, which turn out to be highly accurate. 
In Sec.~\ref{Subsec:MGF} we show topography analysis of the Matsubara Green's 
function $G(k, \beta/2)$ for the Hubbard models with next nearest neighboring (NNN) 
hopping $t'/t=0$ and $t'/t<0$. In Sec.~\ref{Subsec:Pairing} the $d$-wave pairing
responses vs various pinning fields $h_p$ are shown. In Sec.~\ref{StripeOrder} 
we show the spin and charge correlations of $W=4$ doped Hubbard model, which 
indicate a half-filled stripe phase. In Sec.~\ref{ErrorAnalysis} we carefully analyze 
the typical errors of tanTRG at various temperatures.

\begin{figure}[htbp]
\centering
\includegraphics[width = 0.85\linewidth]{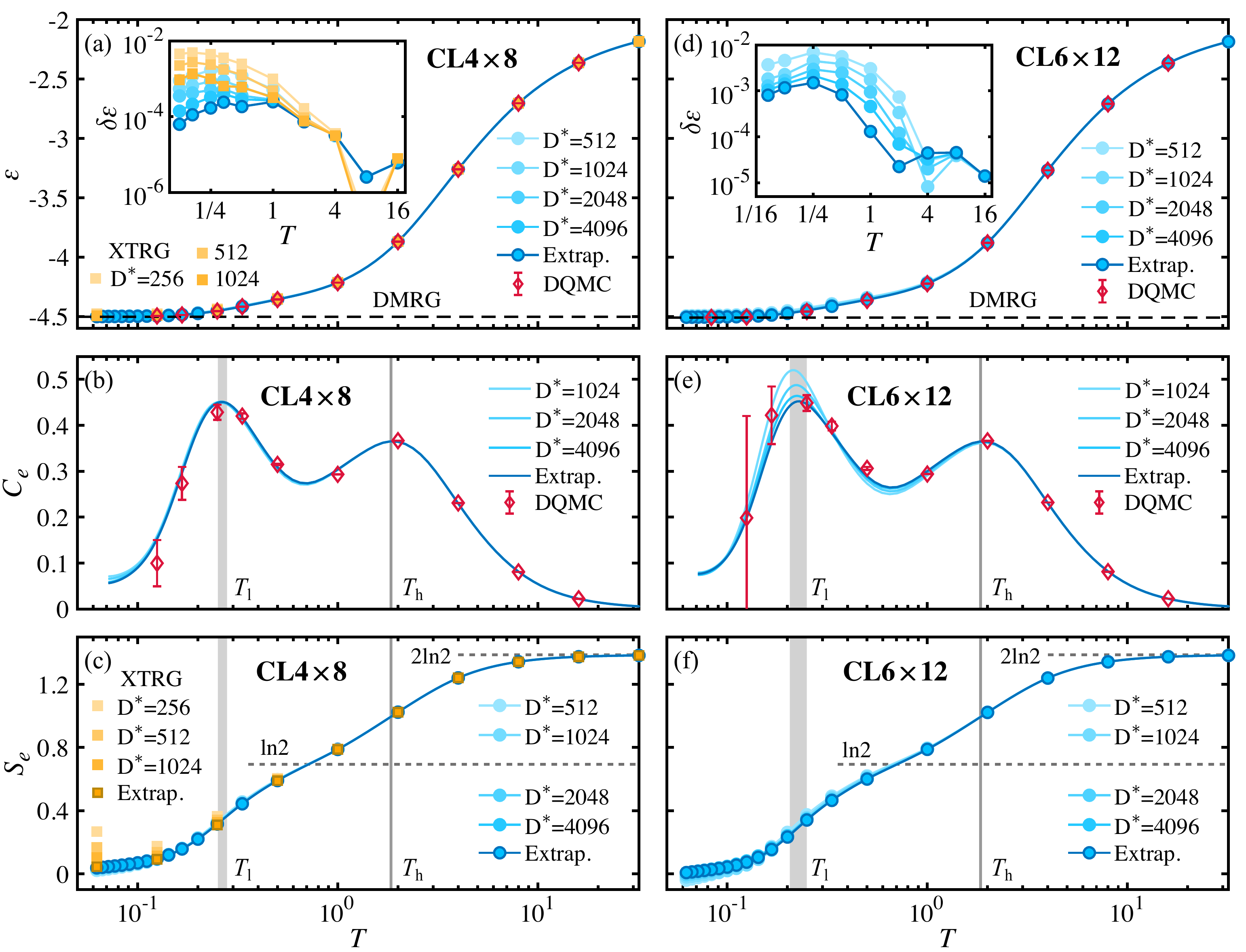}
\caption{Benchmark results on $W=4$ and $W=6$ cylinders. (a-c) show 
the energy $\varepsilon$, specific heat $C_e$, and entropy $S_e$ of CL$4\times8$ 
lattice, and (d-f) show the corresponding results on the CL$6\times12$ lattice. 
The energy $\varepsilon$ and specific heat $C_e$ on both lattices show excellent 
agreements with DQMC calculations. For the CL$4\times8$ lattice, we also 
plot XTRG results of $\varepsilon$ and $S_e$, with retained bond multiplets $D^*=256$, 
512, and 1024, whose extrapolation shows very good agreement with the 
tanTRG and DQMC results. In particular, the insets in (a,d) show the 
relative error of tanTRG results as compared to the standard DQMC data.
} 
\label{Fig_S1}    
\end{figure} 
  
\subsection{Benchmark of Hubbard cylinders at half-filling}
\label{Subsec:Cylinder}
In Fig.~\ref{Fig_S1} we show the tanTRG results of the half-filled 
Hubbard model on $W=4$ and $W=6$ cylinder lattices, denoted as CL4 
and CL6 henceforth. The on-site repulsive interaction is fixed as $U=8$,
and the results are compared to those obtained by determinant quantum 
Monte Carlo (DQMC) and exponential tensor renormalization group 
(XTRG) methods. In Fig.~\ref{Fig_S1}(a,d) we show the energy per 
site $\varepsilon$ vs $T$ with retained bond multiplets up to $D^*=4096$, i.e., 
approximately $D=25,000$ individual states, with which we obtain highly 
accurate tanTRG results on both CL4 and CL6 lattices. The insets in 
Fig.~\ref{Fig_S1}(a,d) plot the relative errors of energy $\delta \varepsilon 
= |\varepsilon-\varepsilon_{\rm DQMC}|/|\varepsilon_{\rm DQMC}|$, 
which are of $O(10^{-4})$ for CL4 and $O(10^{-3})$ for CL6 till the 
lowest temperature $T/t \simeq 0.06$. 
Note they are already comparable to the estimated Trotter errors in practical 
DQMC calculations (c.f., Fig.~\ref{Fig_S11} below). In Fig.~\ref{Fig_S1}(a) 
we also plot the XTRG results and show their relative errors $\delta \varepsilon$ 
in the inset of Fig.~\ref{Fig_S1}(a), from which we see that for the challenging 
Hubbard model and with the same retained bond dimension, the tanTRG 
can produce energy expectation values with higher accuracy than XTRG.

In Fig.~\ref{Fig_S1}(b,e), we show the tanTRG results of electron 
specific heat $C_e$ on $W=4$ and 6 cylinders, and see in both cases again 
two temperature scales. They are higher temperature scale $T_{\rm h}/t\sim 2$ 
that corresponds to the charge degree of freedom and the lower temperature 
scale $T_{\rm l}/t\sim 0.2$ for the spin degrees of freedom. The higher 
temperature scale $T_{\rm h}$ is found to be stable for different system sizes, 
while the lower scale $T_{\rm l}$ is slightly higher for the CL4 than those of 
CL6 and CL8. In Fig.~\ref{Fig_S1}(b,e) we also plot DQMC results of 
$C_e$ and find excellent agreements within error bars for both geometries. 

In Fig.~\ref{Fig_S1}(c), we show tanTRG and XTRG data of thermal entropy 
and find the accordant results decrease from the full entropy $\ln{d}$ ($d=4$ is 
the dimension of local Hilbert space) and exhibit a two-step release at $T_{\rm h}$ 
and $T_{\rm l}$, respectively. In the intermediate-temperature regime between 
the $T_{\rm h}$ and $T_{\rm l}$, there exhibits a shoulder-like structure with a 
fractional entropy $\frac{1}{2} \ln{d}$. 
This can be ascribed to the fact that the two peaks, the higher charge peak
and lower spin one. The former $T_{\rm h}$ corresponds to the appearance 
of local moments, and the antiferromagnetic spin correlations build up near
the lower $T_{\rm l}$ scale.
For the entropy calculations, we see again tanTRG can produce more
accurate results than XTRG with the same bond dimension (and remember that 
tanTRG can retain much larger bond dimension up to $D^*=4096$ vs $1024$ 
for XTRG). In Fig.~\ref{Fig_S1}(f) we plot the tanTRG results of entropy data 
for CL6 case and also witness excellent convergence of $S_e$ vs $D^*$.

\begin{figure}[htbp]
\centering
\includegraphics[width = 0.75\linewidth]{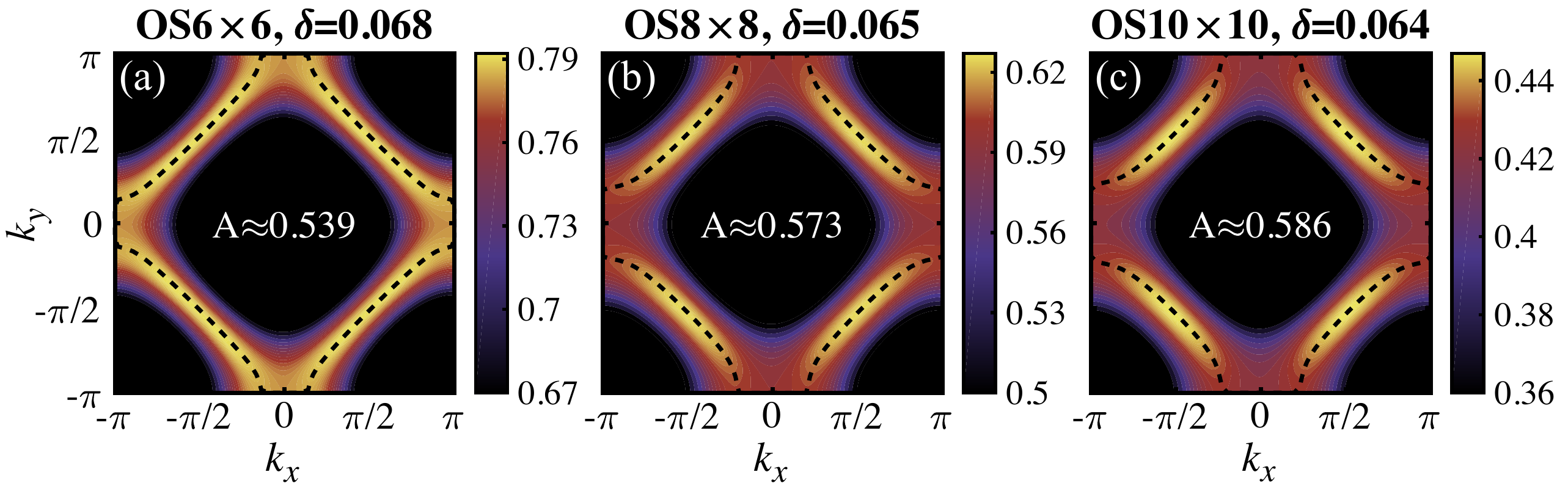}
\caption{(a-c) show the $\beta G(k,\beta/2)$ of doped Hubbard model on $L = 6, 8$ and 
$10$ open square lattices with fixed $\mu = 2$ and $T = 1/3$. The OS$6\times6$ and OS$8\times8$ cases are computed 
with $D^\ast = 2048$ while OS$10\times10$ case with $D^\ast = 4096$.
}
\label{Fig_S2}    
\end{figure} 

\begin{figure}[htbp]
\centering
\includegraphics[width = 0.85\linewidth]{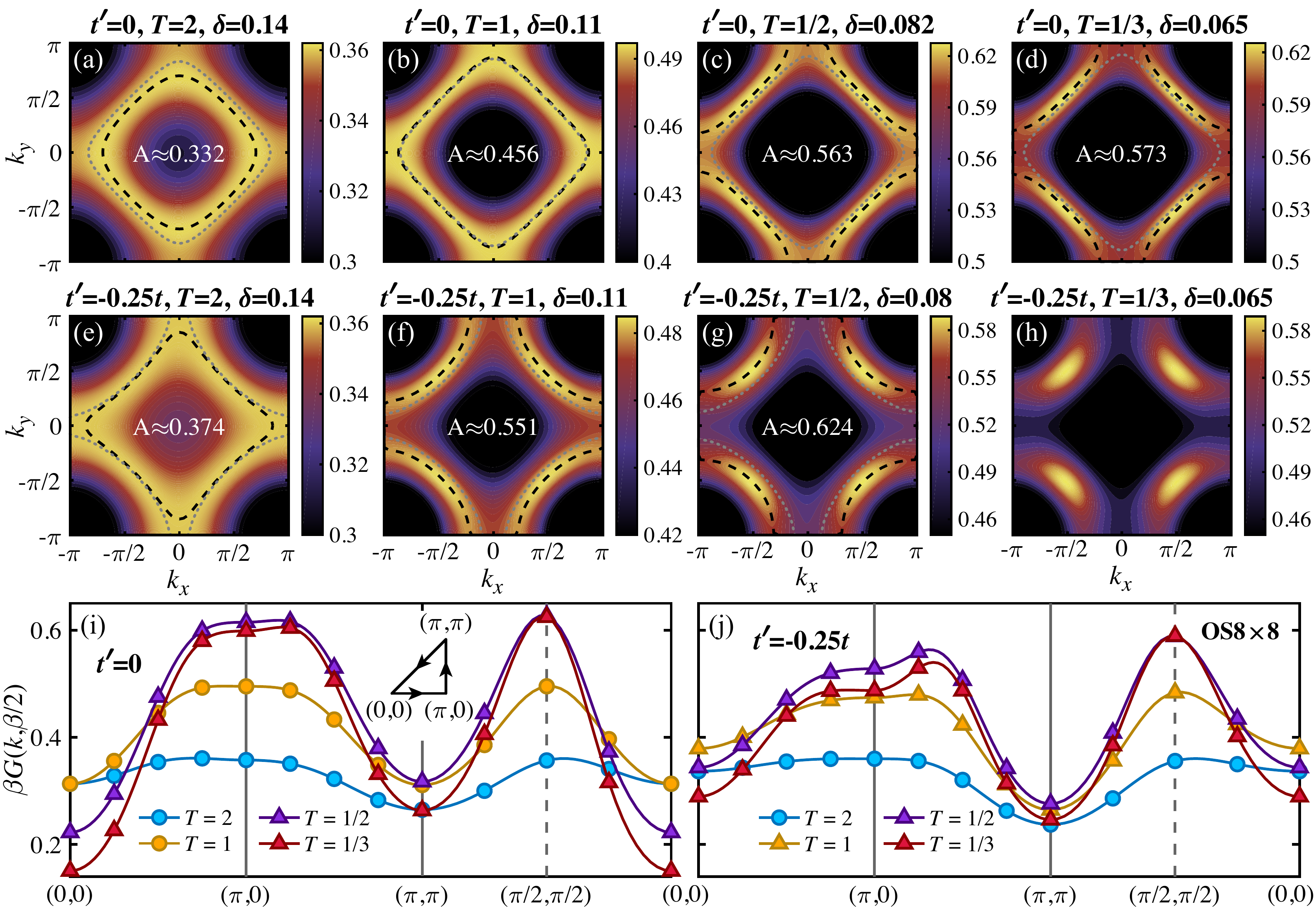}
\caption{(a-d) show the $\beta G(k,\beta/2)$ of OS8$\times$8 doped 
Hubbard model with NNN hopping $t'=0$ and (e-h) for the case with 
$t^\prime/t = -0.25$. The results are obtained with $D^\ast = 2048$, 
$U = 8$ and $\mu = 2$ on temperatures $T = 2, 1, 1/2$ and $1/3$. 
The doping levels vary with $T$ are labeled in the corresponding plots. 
The black dashed lines label the maximal of $\beta G(k,\beta/2)$ and 
provide the estimated Fermi surfaces, where $A$ denotes the area 
enclosed by the Fermi surface. The dotted lines denote the Fermi 
surfaces in the noninteracting limit. (i) and (j) plot the $\beta G(k,\beta/2)$ 
values along the path indicated in the inset of (i), for the $t'=0$ and 
$t'/t=-0.25$ cases, respectively.
}
\label{Fig_S3}    
\end{figure} 

\subsection{Additional results of Matsubara Green's function}
\label{Subsec:MGF}
Now we show the Matsubara Green's function $\beta G(k,\beta/2)$ 
of $t$-$t'$ Hubbard model 
\begin{equation}
H = -t\sum_{\langle i,j\rangle,\sigma} \left(c_{i\sigma}^\dagger c_{j\sigma}^{\phantom{\dagger}} 
+ {\rm H.c.}\right) - t' \sum_{\langle\langle i,j\rangle\rangle,\sigma} \left(c_{i\sigma}^\dagger 
c_{j\sigma}^{\phantom{\dagger}} + {\rm H.c.}\right) + U\sum_{i}n_{i\uparrow}n_{i\downarrow}
-\mu\sum_{i} n_i,
\label{EqS:Hubbard}
\end{equation}
where $t'$ is the hopping between next nearest neighboring (NNN) sites 
$\langle \langle i, j \rangle \rangle$. We perform tanTRG calculations 
of the $t$-$t'$ Hubbard model on square lattice with open boundary condition 
(OS). In Fig.~\ref{Fig_S2}, we compare the results on different system 
sizes, and find the Fermi surfaces become more holelike as the system 
sizes enlarge from OS6$\times$6 to OS10$\times$10. The node-antinode 
feature also becomes more distinguishable as system size increases.

For the OS8$\times$8 geometry, we further compare the results of $t'/t = 0$ 
[Fig.~\ref{Fig_S3}(a-d)] with those of $t'/t=-0.25$ [Fig.~\ref{Fig_S3}(e-h)]. 
In Fig.~\ref{Fig_S3} we find the system with NNN hopping $t'<0$ more favors 
the holelike Fermi surface, and the Lifshitz transition occurs at a clearly higher 
temperature than that of $t'=0$ [c.f., Fig.~\ref{Fig_S3}(f) vs Fig.~\ref{Fig_S3}(c)]. 
More intriguingly, at a low temperature of $T/t=1/3$ [Fig.~\ref{Fig_S3}(h)] 
we find a very prominent node-antinode structure in the $t'/t=-0.25$ case, 
which is also much more prominent than those of $t'=0$ case shown in Fig.
\ref{Fig_S3}(d). 

To see the Matsubara Green's function results more clearly, we show in 
Fig.~\ref{Fig_S3}(i,j) the results along the path indicated in the inset of 
Fig.~\ref{Fig_S3}(i) [i.e., $(0,0) \rightarrow (\pi,0) \rightarrow (\pi,\pi) 
\rightarrow (\pi/2,\pi/2) \rightarrow (0,0)$]. At relatively high temperature, e.g., 
$T/t=2$ and 1, we find similar intensities of spectral weight at $k=(\pi/2, \pi/2)$ 
and $k=(\pi, 0)$ points, for either $t'/t=0$ [Fig.~\ref{Fig_S3}(i)] or $t'/t=-0.25$ 
[Fig.~\ref{Fig_S3}(j)]. However, as temperature further decreases to $T/t=1/2, 
1/3$, the difference in intensity between the $k=(\pi/2, \pi/2)$ and $k=(\pi, 0)$ 
points is visible for $t'/t=0$ and becomes significant for $t'/t=-0.25$.

\subsection{Pairing responses versus pairing fields}
\label{Subsec:Pairing}

\begin{figure}[htbp]
\centering
\includegraphics[width = 0.45\linewidth]{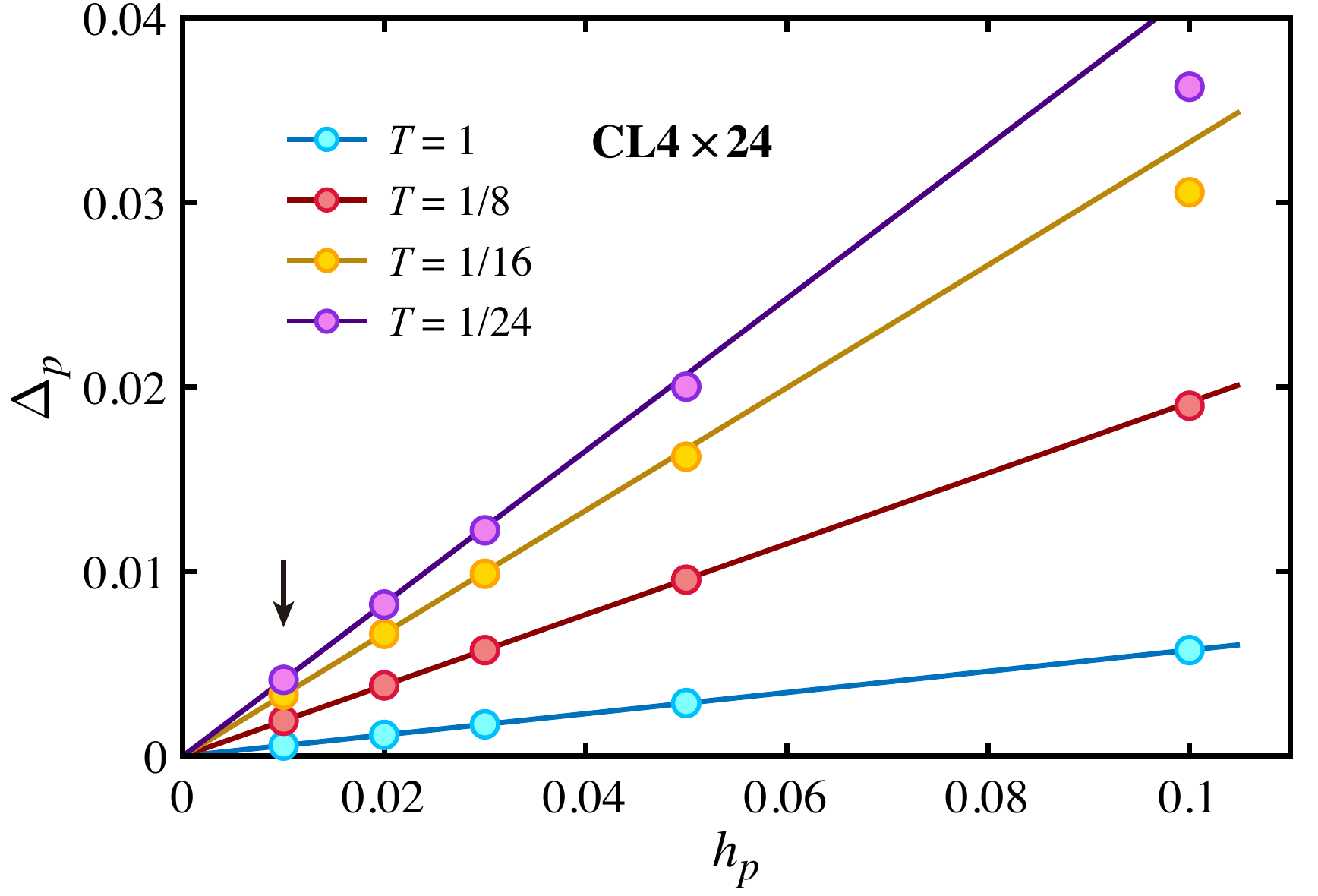}
\caption{The $d$-wave pairing order parameter $\Delta_p$ of the 
CL$4\times24$ Hubbard model with different pairing fields 
$0.01 \leq h_p \leq 0.1$. The results are computed for the 
$\mu = 1.75$ case, with $D^\ast = 1024$ bond multiplets retained. The solid lines connect
the origin and data point $h_p = 0.01$, and extrapolate to the larger $h_p$ regime, from 
which we find the data points with $h_p\leq 0.05$ well fall into a linear response regime.}
\label{Fig_S4}    
\end{figure}

Now we show in Fig.~\ref{Fig_S4} the computed results of pairing order 
parameter $\Delta_{p} = \frac{1}{N_p} \langle \Delta_{\rm tot} \rangle_T$ 
($N_p$ the total bond number) under various paring fields $h_p$. In Fig.~5 
of the main text we have shown the pairing susceptibility $\chi_{\rm SC} 
= \Delta_{p}/h_p$ with a small pinning field $h_p=0.01$. Here in Fig.
\ref{Fig_S4}, we find the chosen $h_p=0.01$ in the main 
text indeed well resides in the linear response regime, and the order 
parameter $\Delta_{p}$ vanishes as $h_p \rightarrow 0$, even for the 
lowest temperature $T/t=1/24$. These observations thus validate that 
the pairing susceptibility $\chi_{\rm SC}$ in Fig.~5 of the main text 
indeed reflects the intrinsic pairing responses of the system. 

\subsection{Spin and charge correlations for the $t'=0$ Hubbard model
\label{StripeOrder}}
\begin{figure}[htbp]
\centering
\includegraphics[width = 0.85\linewidth]{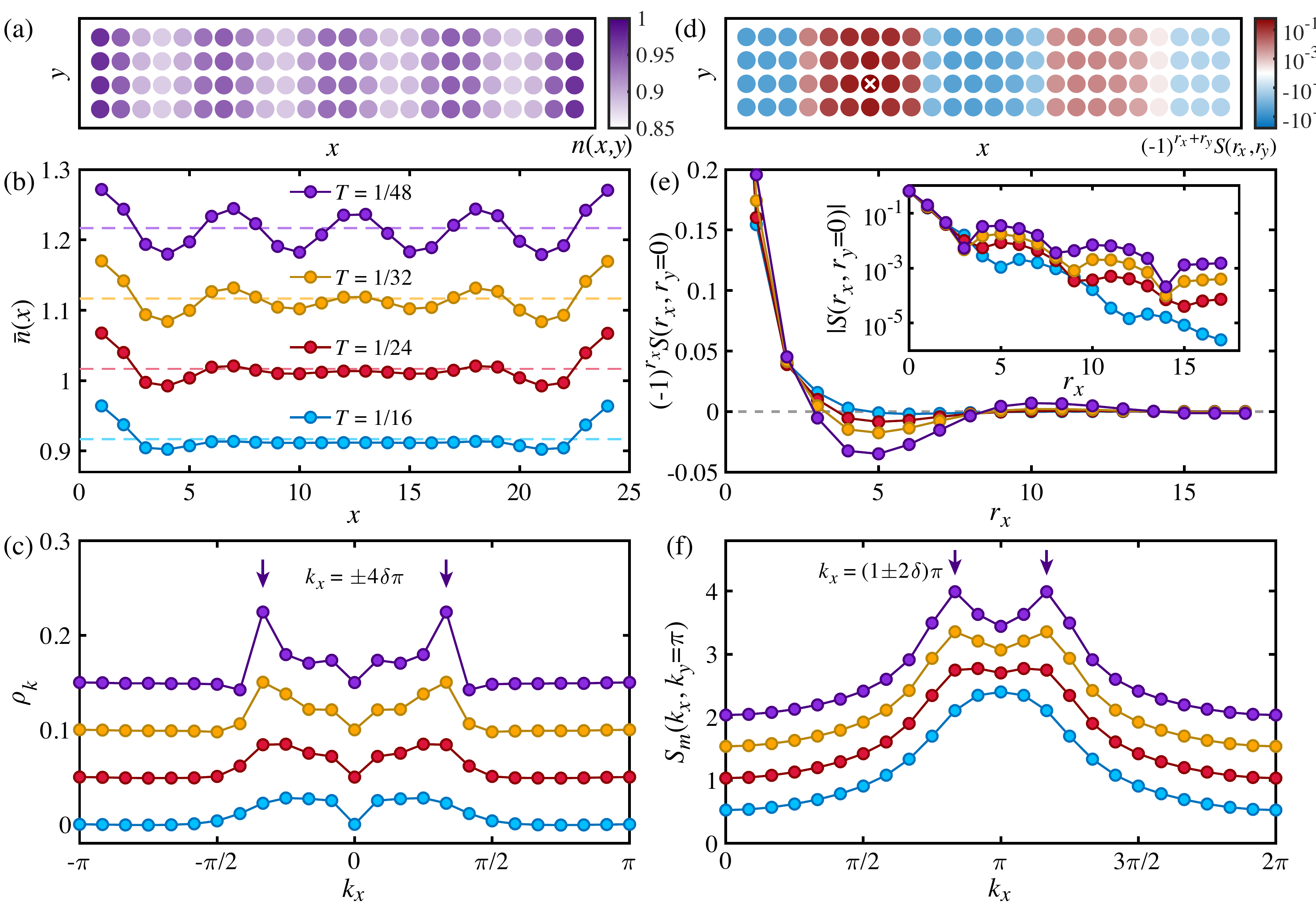}
\caption{(a) shows the electron density $n(x,y)$ computed at low
temperature $T = 1/48$, where the charge stripe can be observed clearly. 
(b) shows the electron density $\bar{n}(x) = \frac{1}{W} \sum_{y=1}^{W} 
n(x,y)$ at various temperatures ($W=4$ denotes the width of the cylinder), 
where the dashed lines denote the average electron density $\bar{n}=1-\delta$. 
(c) shows the Fourier transformation $\rho_k = \frac{1}{\sqrt{L}}\sum_{x = 
1}^{L} e^{-ik_xx}(\bar{n}(x) - \bar{n})$, where $L = 24$ is the length of cylinder, 
and the double peaks located at $k_x = \pm4\delta\pi$ clearly indicate the 
presence of (half-filled) charge stripe in the system. 
(d) shows the spin correlation $(-1)^{r_x+r_y}S(r_x,r_y) = (-1)^{r_x+r_y}
\langle S_{x,y} \cdot S_{x+r_x,y+r_y}\rangle$ at $T = 1/48$, observed 
from the reference point labeled by a cross. 
(e) shows the spin correlations along the length direction, where the 
periodically appearing ``nodes'' in $(-1)^{r_x+r_y}S(r_x,r_y)$ (and 
``dips'' in the absolute value $|S(r_x,r_y = 0)|$ in the inset) correspond 
to the sign change ($\pi$-phase shift) in the short-range SDW. 
(f) shows the spin structure factor $S_m(k) = \frac{1}{LW}\sum_{i,j}e^{-ik
\cdot(r_j-r_i)} \langle S_i\cdot S_j\rangle$, where the peaks located at 
$k_x = (1\pm 2\delta)\pi$ are consistent with SDW observed
on panel (d). 
Note that the curves in panel (b,c,f) are each shifted upwards by, $\Delta 
\bar{n} = 0.1$, $\Delta \rho_k = 0.05$, and $\Delta S_m = 0.5$, for the 
charge density, its Fourier transformation, and spin structure factor results, 
respectively, for the sake of clarity. All the above results are obtained with 
tanTRG by retaining $D^\ast = 2048$ [about 5500 equivalent U(1) 
states] and with a fine tuned chemical potential $\mu 
\approx 1.8$ in the temperature range of interest.
}
\label{Fig_S5}
\end{figure}

In Fig.~\ref{Fig_S5} we show the tanTRG results of charge density distribution 
and spin-spin correlations. The results are computed on CL$4\times24$ geometry 
with $U=8$ and $\delta \approx 1/12$ controlled via a fine tuning of the chemical 
potential $\mu$. 

In Figs.~\ref{Fig_S5}(a-c) we show the electron density distributions on a 
width-4 cylinder. A charge stripe pattern is apparent at the lowest temperature 
$T=1/48$ as visualized in Fig.~\ref{Fig_S5}(a), with its temperature evolution 
indicated in Fig.~\ref{Fig_S5}(b). The charge stripe can be already observed 
at $T=1/24$, and becomes quite prominent for $T\leq 1/32$. This can also be 
clearly recognized in the Fourier transformation of electron density, $\rho_k$, 
shown in Fig.~\ref{Fig_S5}(c). From the results in Fig.~\ref{Fig_S5}(a-c), we 
observe a half-filled stripe with wave length $\lambda_{\rm CDW} = 6 \approx 
1/(2\delta)$ at sufficiently low temperature. Our calculations are consistent with 
previous results in Refs.~\cite{Jiang2022PNAS,Hao2022PRR}.

Regarding the spin-spin correlations, we also find a spin density wave (SDW) 
in Figs.~\ref{Fig_S5}(d-f). From a reference point in the bulk [labeled by a cross 
in Fig.~\ref{Fig_S5}(d)], we find the correlation $(-1)^{r_x+r_y} S(r_x, r_y)$ 
changes its sign every 6 sites along the length direction, i.e., a $\pi$-phase 
shift occurs for $\lambda_{\rm SDW}/2 = 6$ that equals $\lambda_{\rm CDW}$. 
Besides, as temperature lowers the spin-spin correlation strengthens 
[see Fig.~\ref{Fig_S5}(e)] and the structure factors exhibit peaks at $k_x = 
(1\pm 2\delta) \pi$ [Fig.~\ref{Fig_S5}(f)]. Our findings are consistent with 
the $W=4$ results reported in Refs.~\cite{Hao2022PRR} and \cite{Xiao2023PRX}.

\begin{figure}[htbp]
\centering
\includegraphics[width = 0.75\linewidth]{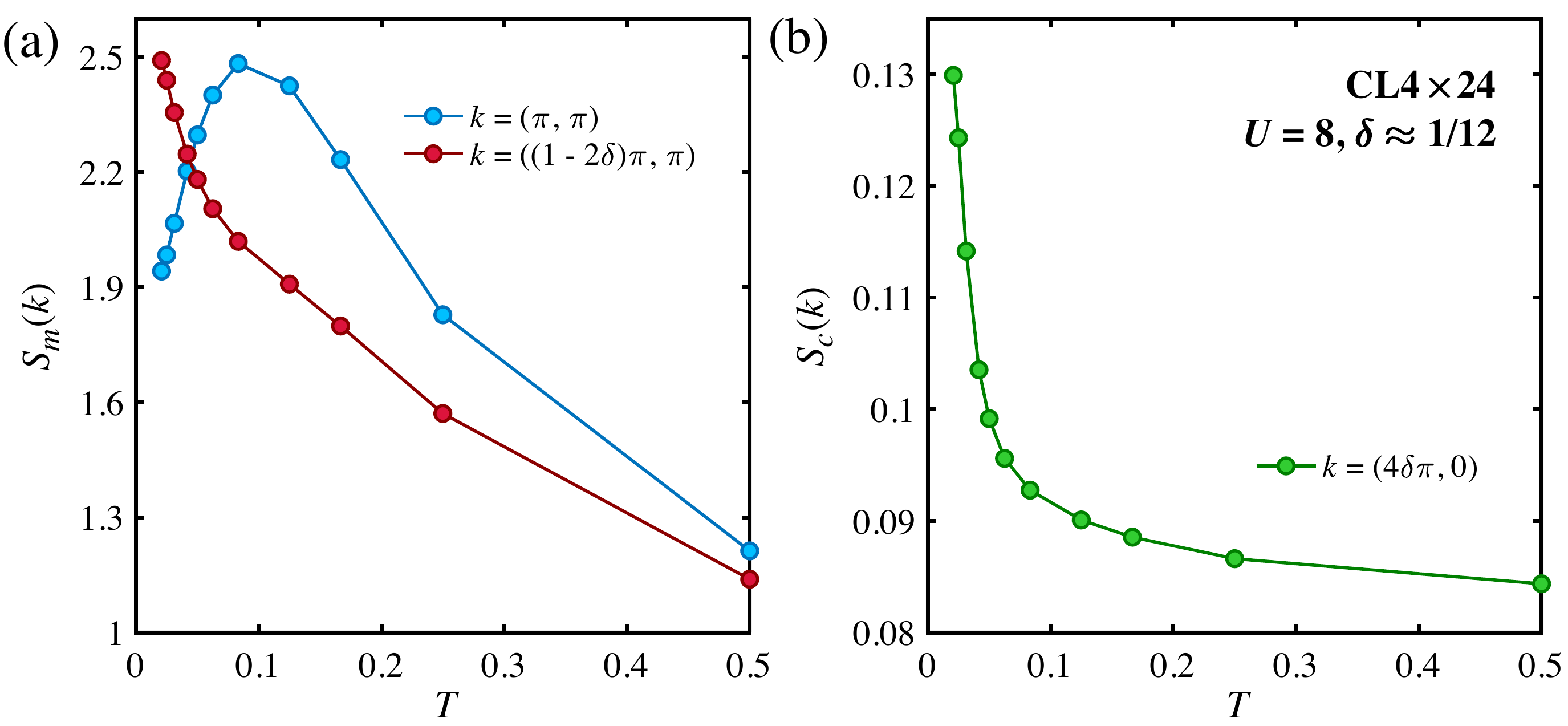}
\caption{(a,b) show the tanTRG results on CL$4\times24$ lattice, 
with $U = 8$ and approximately fixed doping $\delta \approx 1/12$ 
(controlled by fine tuning $\mu$ in the grand canonical ensemble simulations). 
The tanTRG results are obtained with $D^\ast = 2048$ [about 5500 
equivalent U(1) states].}
\label{Fig_S6}
\end{figure}

In Fig.~\ref{Fig_S5} we have already shown the results of half-filled stripe 
obtained on the width-4 cylinder that are in agreement with the minimally 
entangled typical thermal states (METTS) results~\cite{Wietek2021PRX}. 
In Fig.~\ref{Fig_S6}, we further show that the spin and charge structure factors 
obtained by tanTRG exhibit similar behaviors to METTS results, though 
with different on-site interactions and dopings, i.e., $U = 8$ and $\delta \approx 
1/12$ in tanTRG vs $U = 10$ and $\delta=1/16$ in METTS. The presence 
of a hump in $S_m(\pi, \pi)$ at intermediate temperature and the ever increasing 
$S{_m}((1-2\delta)\pi, \pi)$ curve with a shoulder structure are reproduced 
in our tanTRG calculations. On the other hand, the charge structure factor 
at $k=(4 \delta \pi, 0)$ increases rapidly as temperature decreases 
[c.f., Fig.~\ref{Fig_S6}(b)], also in qualitative agreement with METTS.

Last, it is worthwhile discussing the advantages and disadvantages 
of the two methods, tanTRG and METTS. They represent two 
typical tensor-network approaches, MPO-based (purification) vs 
MPS-based (with Monte Carlo samplings), for finite temperature simulations. 
Unlike current METTS calculations that use Trotter decomposition 
and swap gates (at least initially) and require separate runs for each 
temperature point, tanTRG generates all temperature points in one 
run and is free of statistical errors. It is straightforward to implement 
Abelian and non-Abelian symmetries in tanTRG, which further improves 
its efficiency and enables wider systems (width-8 vs width-4 cylinders) 
to be calculated. As the MPO representation of density operator is available, 
it is more convenient to compute certain quantities like the imaginary-time 
correlations, pairing susceptibilities, etc, in tanTRG. On the other hand,
METTS works with the canonical ensemble, and thus it is easier
to control the particle number in the calculations. Moreover, METTS is well
suited for multi-core parallel computing as it uses Monte Carlo samplings in 
its calculations.

\subsection{Error analysis of tanTRG\label{ErrorAnalysis}}
\begin{figure}[htbp]
\centering
\includegraphics[width = 0.85\linewidth]{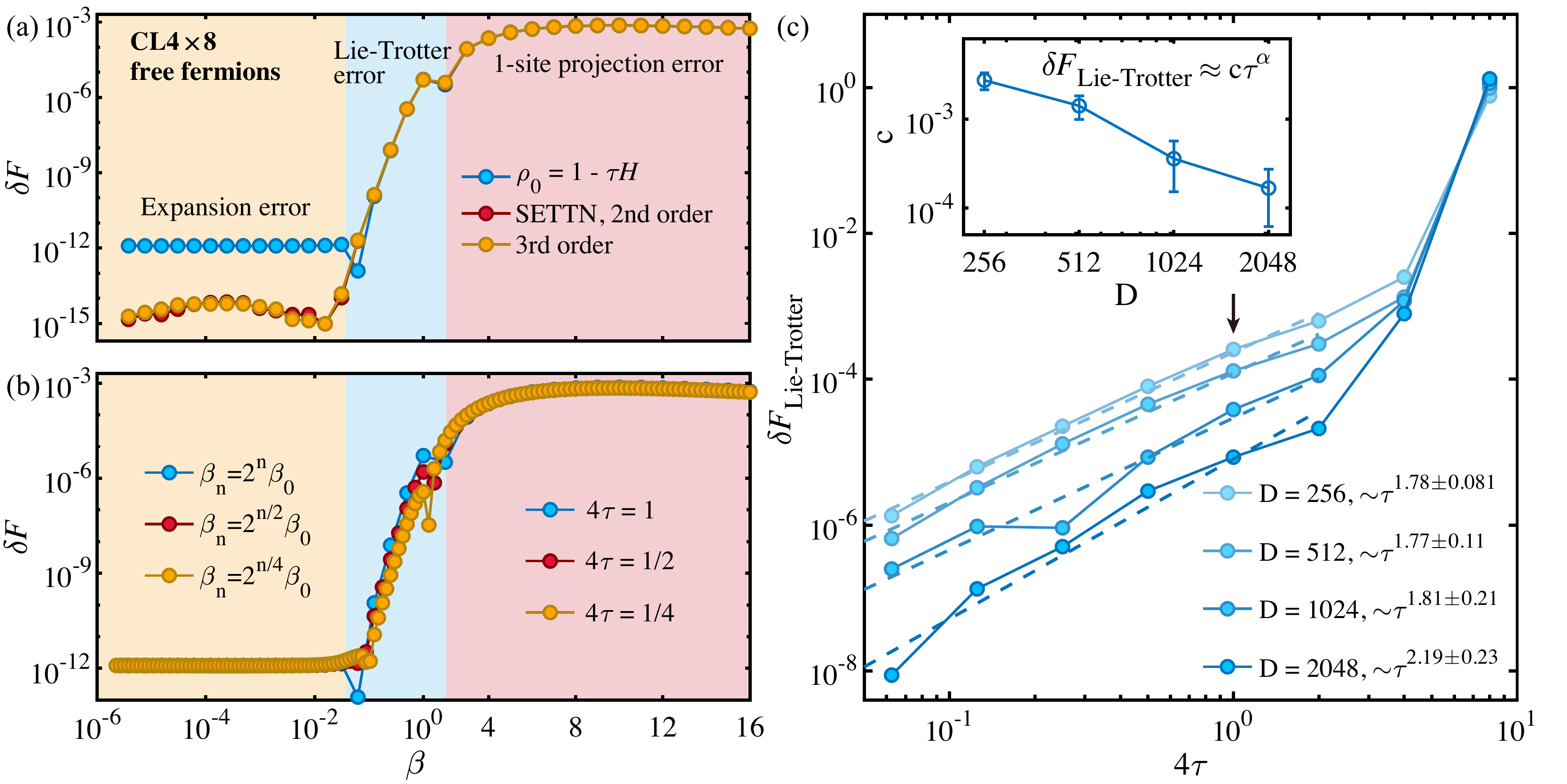}
\caption{Relative errors in free energy of CL$4\times8$ free fermions. 
(a) The SETTN initialization reduces the expansion error at high temperatures 
down to the machine precision. (b) compares the Lie-Trotter errors with 
different temperature grids, more specifically $\beta_n = 2^n\beta_0,  
2^{n/2}\beta_0$, and $2^{n/4} \beta_0$ used in exponential cooling and 
$4\tau = 1, 1/2$ and $1/4$ in the linear cooling. The errors curves of different 
grids coincide, except for in the intermediate-temperature regime. Moreover, 
when decreasing the step length, the second (lower-$T$) cancellation point 
moves towards 
higher temperature since the Lie-Trotter error decreases while the 1-site 
projection error is independent on step length. Both (a) and (b) are computed by 
tanTRG with a fixed $D = 1024$. (c) estimates the Lie-Trotter errors by 
$\delta F_{\rm Lie-Trotter} = |F - F_0|/|F_0|$ at a fixed (inverse) temperature
$\beta = 9$. In particular, $F_0$ is also obtain by tanTRG with a sufficiently 
small $\tau = 1/128$, and all calculations are evolved from the same density 
operator $\rho(\beta \equiv 1)$. $\delta F_{\rm Lie-Trotter}$ results are found 
fall into a $c \tau^\alpha$ behavior with $\alpha\approx 2$ and the prefactor 
$c$ controlled by $D$ as shown in the inset. The black arrow denotes the step 
length we used throughout in the main text, which corresponds to negligible 
Lie-Trotter error as compared to the dominant 1-site projection error.
}
\label{Fig_S7}    
\end{figure}

In tanTRG, there exist different types of errors. The majors are three, 
expansion error in the initialization, Lie-Trotter error, and the projection error. 
In Fig.~1(c) of the main text, we observed two dips, i.e., cancellation points 
that naturally separates three regimes where the three types of errors dominate. 
In the high-temperature regime, the 1st-order expansion $\rho_0 = 1-\tau_0 H$ 
is the main resource of error, which can be suppressed by the Taylor expansion 
to higher order, namely, $\rho_0 = \sum_{n=0}^{N_{c}} \frac{(-\tau_0 H)^n}{n!}$ 
expanded up to a sufficiently large $N_{c}$, with the techniques developed 
in series-expansion thermal tensor network (SETTN) approach~\cite{SETTN}. 
As shown in Fig.~\ref{Fig_S7}(a), the relative error $\delta F$ for $\beta
{\lesssim} 10^{-2}$ can be suppressed to machine precision, i.e., 
$\sim 10^{-15}$ with the SETTN initialization.

In the intermediate-temperature regime marked by blue color in Fig.
\ref{Fig_S7}(a,b), the Lie-Trotter splitting becomes the main source 
of computational errors. In Fig.~\ref{Fig_S7}(b), we find the Lie-Trotter 
errors in this intermediate regime decrease as step length $\tau$ is reduced. 
Since that the sign of Lie-Trotter errors turns out to be different with expansion 
error, there exists a cancellation point at $\beta \sim 10^{-2}$ (i.e., a dip). 
Note that the location of this first dip can be pushed to high-temperature 
side when SETTN technique is used in the initialization [c.f., Fig.
\ref{Fig_S7}(a) where the cancellation point becomes virtually invisible 
due to the negligible expansion error with SETTN intialization].

Finally, at low temperature $\beta \gtrsim 1$ the 1-site projection errors, 
which reflect the expression capability of MPO representation with 
given bond dimension $D$, become sufficiently large and dominate over 
all other type of errors after the second dip.

In Fig.~\ref{Fig_S7}(c) we show the scaling analysis of Lie-Trotter error at a 
fixed (inverse) temperature $\beta=9$, where the relative errors $\delta F_{\rm 
Lie-Trotter}$ are found to scale as $c \tau^\alpha$ with $\alpha \approx 
2$. The fitted prefactor $c$ is plotted in the inset, from which we find the prefactor, 
and thus the overall Lie-Trotter errors, can be well controlled by the bond dimension 
$D$. This surprising finding is in stark difference to the Trotter-Suzuki decomposition 
errors in other thermal TN methods, where the Trotter error is independent of retained 
MPO bond dimension~\cite{Li2011}. 

Moreover, the 2-site projection errors in the initial stage of MPO cooling, 
which is sometimes problematic for MPS-based TDVP calculations
(e.g., when starting from a direct product state), are largely absent in 
tanTRG. A possible reason is that the additional physical indices in 
MPO introduce more variational parameters, bearing some similarity 
to the role of subspace expansion in MPS for mitigating the large 
initial 2-site projection errors~\cite{MingruYang2020,BondAdapt1tdvp2021}.

\section{Derivation of Tangent Space Approach for Thermal Tensor Networks}
\label{Sec:TSA4TTN}
Here we provide details in deriving the flow equation of density operator 
$\rho$ in the imaginary-time evolution, as well as technical details of the 
algorithms involved in the main text. For the sake of simplicity, we firstly 
introduce the Choi transformation that maps the matrix product operator 
(MPO) $\rho$ to a matrix product state (MPS) $\kett{\rho}$ in Sec.~\ref{Sec_Choi}, 
which thus allows us to follow the standard derivation of time dependent 
variational principle (TDVP) approach~\cite{TDVP2011,TDVP2016}, with 
some adaptation to the super MPS mapped from MPO. We present a short 
introduction to the MPS manifold and tangent vectors in Sec.~\ref{Sec_MPS}, 
and the orthogonal condition~\cite{MPSManifold2014,GeometryTDVP2020} 
in Sec.~\ref{Sec_Tangent}. After that, we derive the flow equation induced 
by the optimal tangent vectors in Sec.~\ref{Sec_OptimalB}, and then introduce 
the splitting method for integration together with the Lie-Trotter error analysis 
in Sec.~\ref{Sec_LieTrotter}. The single-site 1-TDVP integrator is detailed in 
Sec.~\ref{Sec_procedure}, and two-site 2-TDVP in Sec.~\ref{Sec_2TDVP}. 
The latter allows the bond dimension to increase and is very useful in the 
imaginary-time evolution of $\rho$. Another useful technique, Lanczos exponential 
method used for updating local tensors, is present in Sec.~\ref{Sec_Lanczos}. 
Lastly, bearing in mind that the super MPS language, essentially equivalent to 
MPO representation, is introduced here merely to facilitates the derivatives of 
flow equation in the main text.

\subsection{Choi transformation \label{Sec_Choi}}
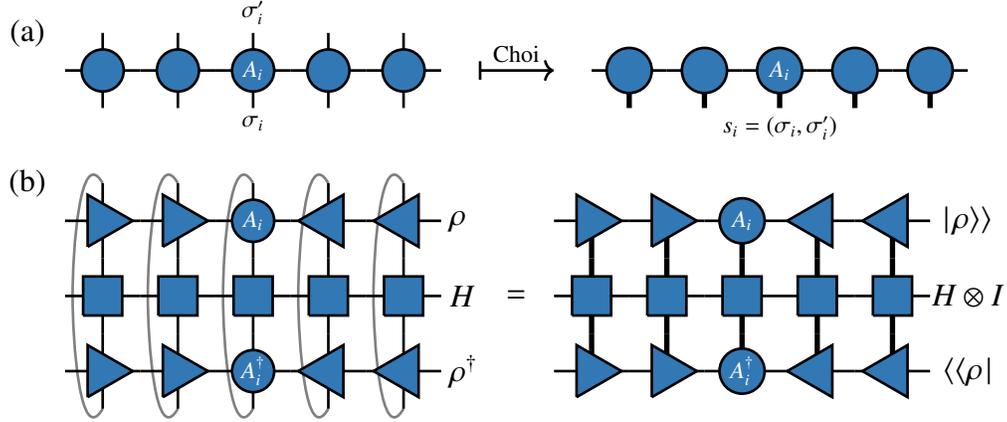
\begin{figure}[htbp]
    \centering
    \begin{tikzpicture}[scale = 1]
        \node at (0, 0.5) {\large (a)};
        \MPOCirc{1}{0}{}
        \MPOCirc{2}{0}{}
        \MPOCirc{3}{0}{$A_i$}
        \node [below] at (3, -0.5) {$\sigma_i$};
        \node [above] at (3, 0.5) {$\sigma_i^\prime$};
        \MPOCirc{4}{0}{}
        \MPOCirc{5}{0}{}

        \draw [|->, line width = 1] (6, 0) -- (7, 0);
        \node [above] at (6.5, 0) {Choi};

        \MPSCirc[2]{8}{0}{}
        \MPSCirc[2]{9}{0}{}
        \MPSCirc[2]{10}{0}{$A_i$}
        \node [below] at (10, -0.5) {$s_i = (\sigma_i,\sigma_i^\prime)$};
        \MPSCirc[2]{11}{0}{}
        \MPSCirc[2]{12}{0}{}

        \node at (0, -1.5) {\large (b)};
        \draw [gray, line width = 1] (1, -1.5) [out = 120, in = 90] to (0.6, -3) [out = -90, in = -120] to (1, -4.5);
        \draw [gray, line width = 1] (2, -1.5) [out = 120, in = 90] to (1.6, -3) [out = -90, in = -120] to (2, -4.5);
        \draw [gray, line width = 1] (3, -1.5) [out = 120, in = 90] to (2.6, -3) [out = -90, in = -120] to (3, -4.5);
        \draw [gray, line width = 1] (4, -1.5) [out = 120, in = 90] to (3.6, -3) [out = -90, in = -120] to (4, -4.5);
        \draw [gray, line width = 1] (5, -1.5) [out = 120, in = 90] to (4.6, -3) [out = -90, in = -120] to (5, -4.5);
        \MPOTriaL{1}{-2}{}
        \MPOSqua[1]{1}{-3}{}
        \MPOTriaL{1}{-4}{}
        \MPOTriaL{2}{-2}{}
        \MPOSqua[1]{2}{-3}{}
        \MPOTriaL{2}{-4}{}
        \MPOCirc{3}{-2}{$A_i$}
        \MPOSqua[1]{3}{-3}{}
        \MPOCirc{3}{-4}{$A_i^\dagger$}
        \MPOTriaR{4}{-2}{}
        \MPOSqua[1]{4}{-3}{}
        \MPOTriaR{4}{-4}{}
        \MPOTriaR{5}{-2}{}
        \MPOSqua[1]{5}{-3}{}
        \MPOTriaR{5}{-4}{}
        \node [right] at (5.5, -2) {\large $\rho$};
        \node [right] at (5.5, -3) {\large $H$};
        \node [right] at (5.5, -4) {\large $\rho^\dagger$};

        \node at (6.5, -3) {\large $=$};

        \MPSTriaL[2]{7.5}{-2}{}
        \MPOSqua[2]{7.5}{-3}{}
        \AMPSTriaL[2]{7.5}{-4}{}
        \MPSTriaL[2]{8.5}{-2}{}
        \MPOSqua[2]{8.5}{-3}{}
        \AMPSTriaL[2]{8.5}{-4}{}
        \MPSCirc[2]{9.5}{-2}{$A_i$}
        \MPOSqua[2]{9.5}{-3}{}
        \AMPSCirc[2]{9.5}{-4}{$A_i^\dagger$}
        \MPSTriaR[2]{10.5}{-2}{}
        \MPOSqua[2]{10.5}{-3}{}
        \AMPSTriaR[2]{10.5}{-4}{}
        \MPSTriaR[2]{11.5}{-2}{}
        \MPOSqua[2]{11.5}{-3}{}
        \AMPSTriaR[2]{11.5}{-4}{}
        \node at (12.5, -2) {\large $\kett{\rho}$};
        \node at (12.5, -3) {\large $H\otimes I$};
        \node at (12.5, -4) {\large $\bbra{\rho}$};

\end{tikzpicture}
\caption{(a) shows the mapping between MPO and (super) MPS via the 
Choi transformation. Note that we set the physical index a \textit{bold} line
to emphasize that it a combined index $s_i \equiv (\sigma_i, \sigma'_i)$. 
(b) depicts the equivalence between the tensor network representations 
of $\Tr{\rho^\dagger H\rho}$ (left) and $\bbra{\rho}H\otimes I\kett{\rho}$ 
(right). By substituting the rank-4 tensor $A_i$ in MPO with rank-3 local tensor
$A_i$ (with a bold index) in MPS, we arrive at the right-hand-side of the equation.
Note the identity $I$ at the right-hand-side of the equation will be 
skipped henceforth for the sake of simplicity.
}
\label{Fig_Choi}
\end{figure}

We denote the Hilbert space as $\mathcal{H}$ and the space of bounded linear 
operators on $\mathcal{H}$ as $B(\mathcal{H})$, and the Choi transformation is
defined as follows
\begin{equation}
\begin{aligned}
B(\mathcal{H}) & \, \rightarrow \, \mathcal{H}\otimes\mathcal{H}\\
\rho = \sum_{ij}\rho_{ij}\ket{i}\bra{j} & \, \mapsto \, \kett{\rho} \equiv \sum_{ij}\rho_{ij}\ket{i}\otimes\ket{j},
\end{aligned}
\end{equation}
which represents an isomorphism between Hilbert spaces. Note that the
Choi transformation depends on the specific basis $\{\ket{i}\}$ that we 
choose. In particular, an MPO can be mapped into a (super) MPS via the
Choi transformation as shown in Fig.~\ref{Fig_Choi}(a). Meanwhile, 
for a linear super operator $\mathcal{L}$ acting on $\rho$, represented 
as $\rho \mapsto A\rho B$, the Choi transformation is as follows
\begin{equation}
\mathcal{L}\rho = A\rho B \, \mapsto \, A\otimes B^T\kett{\rho},
\end{equation}
which satisfies the commutative diagram
\begin{equation}
    \begin{tikzcd}
        B(\mathcal{H}) 
            \arrow[r, "\text{Choi}"]
            \arrow[d, "\mathcal{L}"'] 
        & {\mathcal{H}\otimes\mathcal{H}}
            \arrow[d, "A\otimes B^T"] 
        \\
        B(\mathcal{H})
            \arrow[r, , "\text{Choi}"] 
        & {\mathcal{H}\otimes\mathcal{H}}.
    \end{tikzcd}
\end{equation}

With this transformation, we map the imaginary-time evolution equation 
$d\rho/d\beta = -H\rho$ to an equation of state, i.e.,
\begin{equation}
   \frac{d}{d\beta}\kett{\rho} = - H\otimes I \kett{\rho}.
\end{equation} 
Although in principle we need to evolve the super MPS $\kett{\rho}$ 
according to the super operator $H \otimes I$, it can be simplified by 
taking virtual of the fact {$e^{H \otimes I} = e^{H}\otimes I$}, i.e., the 
identity operator $I$ acting on one-half of physical indices is trivial and 
can thus be safely neglected.

To obtain the expectation values, we need to compute the trace with
density operators, which can be represented as an inner product of 
super MPS, e.g., the energy expectation can be computed as
\begin{equation}
     E(\beta) = \frac{\Tr{\rho^\dagger(\beta/2) \,H\, \rho(\beta/2)}}{\Tr{\rho^\dagger(\beta/2)  \rho(\beta/2)}} = \frac{ \bbra{\rho(\beta/2)}H\otimes I\kett{\rho(\beta/2)}}{\bbra{\rho(\beta/2)}\rho(\beta/2)\rangle\rangle},
\end{equation} 
with the corresponding tensor network representations shown in 
Fig.~\ref{Fig_Choi}(b). Note we have used 
a bilayer trick in the Gibbs operator representation~\cite{BilayerLTRG}, 
i.e., $\rho(\beta) = \rho(\beta/2) \, \rho^\dagger(\beta/2)$. 

The second $\mathcal{H}$ of the Hilbert space $\mathcal{H}\otimes
\mathcal{H}$ is an auxiliary system (B), and thus the partial trace gives 
the density matrix of the physical system (A)
\begin{equation}
\begin{aligned}
{\rm Tr}_B\kett{\rho(\beta/2)}\bbra{\rho(\beta/2)} =& \sum_{ijkl}\rho_{ij}\rho^\ast_{kl} 
{\rm Tr}_B\ket{i}_{A}\ket{j}_B\bra{k}_A\bra{l}_B = \sum_{ijkl}\rho_{ij}
\rho^\ast_{kl}\delta_{jl}\ket{i}_A \bra{k}_A\\
=& \sum_{ijk}\rho_{ij}\rho_{kj}^\ast \ket{i}_A\bra{k}_A = \sum_{ik}
\left(\rho(\beta/2)\rho^\dagger(\beta/2)\right)_{ik}\ket{i}_A\bra{k}_A \\
=& \rho(\beta/2)\rho^\dagger(\beta/2) = \rho(\beta),
\end{aligned}
\end{equation}
which means the supervector $\kett{\rho(\beta/2)}$ actually represents 
a purification of the Gibbs operator $\rho(\beta)$.

\subsection{MPS/MPO manifold and gauge redundancy \label{Sec_MPS}}
Through the Choi transformation, we can map the MPO to a super MPS. 
The latter can generically be represented as 
\begin{equation}
\ket{\Psi(A)} = \sum_{\{s_i\}}\Tr{A_1^{s_1}\cdots A_N^{s_N}}\ket{s_1\cdots s_N}, 
\end{equation}
where each $A_i$ is a rank-3 tensor with a bold physical index $s_i = 1,\cdots 
\tilde{d}_i$, with $\tilde{d}_i = d_i^2$, dimension of local Hilbert space squared
and two bond indices running over $1,\cdots, D_{i-1}$ and $1,\cdots, D_i$,
for $(i-1)$-th and $i$-th bonds, respectively.
\begin{equation}
A = (A_1,\cdots,A_N) \in \bigoplus_{i=1}^{N}\mathbb{C}^{D_{i-1}D_{i}\tilde{d}_i}
\equiv \mathbb{A}
\end{equation}
denotes the parameters of MPS. An MPS is full-rank if there are exactly 
$D_i$ non-zero singular values for each bond. A subset comprised of all 
full-rank MPSs in $\mathbb{A}$ forms a manifold $\mathcal{A}$. Note 
that the MPS representation always has gauge redundancy, i.e., two
MPSs may be essentially equivalent up to a gauge transformation
\begin{equation}
    A_i^{s_i}\,{\mapsto}\, G_{i-1}^{-1}A_i^{s_i}G_{i},
\end{equation}
where $(G_1,\cdots,G_{N-1}) \in \mathcal{G} = \prod_{i = 1}^{N-1}
\mathrm{GL}(D_i, \mathbb{C})$ is the gauge group acting on the bond space 
and $G_0 = G_N \equiv 1\in \mathrm{GL}(1, \mathbb{C})$ specially. With this
gauge group action, $\mathcal{A}$ forms a principal bundle with base manifold 
$\mathcal{M} = \mathcal{A}/\mathcal{G}$, and $\mathcal{M}$ is right the MPS 
manifold, a regular submanifold embedding to the Hilbert space $\mathcal{H}$. 

To (partially) fix the gauge, one can introduce the canonical form of MPS. For example, the canonical condition of local tensor $A_i$ reads 
\begin{equation}
    \left\{
    \begin{aligned}
        &\sum_{s_i}{A_i^{s_i}}^\dagger A_i^{s_i} = cI_{D_i},\quad &(\text{left})\\
        &\sum_{s_i}A_i^{s_i}{A_i^{s_i}}^\dagger = cI_{D_{i-1}},\quad &(\text{right})\\
    \end{aligned}
    \right.
\end{equation}
where $c$ is a constant normalization factor and $I_{D_{i}}$ the 
$D_i \times D_i$ identity matrix.
One can gauge the local tensors to left- or right-canonical form, denoted as 
$A^L$ or $A^R$ respectively. Below we take the gauge convention 
of MPS as follows: it has a site called canonical center, where all the 
local tensors on its left are left-canonical and right-canonical on the right.

\subsection{Tangent vectors of MPS manifold \label{Sec_Tangent}}
A tangent vector on the manifold forms like a summation of $N$ MPSs 
with the $i$-th local tensor $A_i$ varied, as observed by
\begin{equation}
\frac{d}{dt}\ket{\Psi(A)} = \sum_{i = 1}^N \frac{\partial \ket{\Psi(A)}}{\partial A_i}\cdot \frac{d A_i}{d t} 
= \sum_{i = 1}^N \sum_{\{s_j\}}\Tr{{A}_1^{s_1} \cdots 
{{A}_{i-1}^{s_{i-1}} \frac{d A_i^{s_i}}{dt} {A}_{i+1}^{s_{i+1}} \cdots {A}_N^{s_N}}}\ket{s_1\cdots s_N}.
\end{equation}
Following the mixed gauge convention of the MPS tangent vectors, we 
place each varied tensor at the canonical center of the MPS, and arrive 
at a surjective tangent map $\Phi: T\mathcal{A} \rightarrow T\mathcal{M}$, i.e.,
\begin{equation}
\ket{\Phi_A(B)} = \sum_{i = 1}^N \ket{\Phi_A^i(B_i)}
= \sum_{i = 1}^N\sum_{\{s_j\}}\Tr{{(A^L)}_1^{s_1} \cdots 
{{(A^L)}_{i-1}^{s_{i-1}} \,\, B_{i}^{s_i} \,\, {(A^R)}_{i+1}^{s_{i+1}} \cdots 
{(A^R)}_N^{s_N}}}\ket{s_1\cdots s_N},
\label{Eq_tangentMap}
\end{equation}
where the tangent bundle $T\mathcal{A}$ is parameterized as
$(A,B) = (A_1, \cdots, A_N, B_1,\cdots, B_N)$. According to this 
tangent map, any flow equation 
\begin{equation}
    \frac{d}{dt}\ket{\Psi(A)} = \ket{\Phi_A(B)} \in T_{\ket{\Psi(A)}}\mathcal{M}
    \label{EqSM:FlowEq}
\end{equation}
on the MPS manifold can be pulled back to the flow equation $dA/dt = B$ 
in the parameter space. Note that flow equation Eq.~(\ref{EqSM:FlowEq}) 
describes the evolution of state $\ket{\Phi_A(B)}$, and we 
need the flow equation for the local tensors that is convenient 
to deal with in the algorithmic level. 
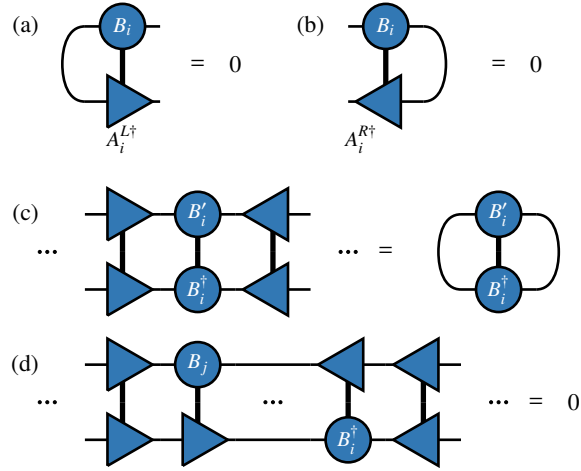
\begin{figure}[htbp]
    \centering
    \begin{tikzpicture}[scale = 1]
        \node [left] at (0, 1.5) {(a)};
        \BEnvL{0.5}{0.5}{1.5}{}
        \MPSCirc{1}{1.5}{\footnotesize $B_i$}
        \AMPSTriaL{1}{0.5}{}
        \node at (1, 0) {\footnotesize $A_i^{L\dagger}$};
        \node at (2, 1) {$=$};
        \node at (2.5, 1) {$0$};

        \node at (3.5, 1.5) {(b)};
        \MPSCirc{4.5}{1.5}{\footnotesize$B_i$}
        \AMPSTriaR{4.5}{0.5}{}
        \node [left] at (4.5, 0) {\footnotesize$A_i^{R\dagger}$};
        \BEnvR{5}{0.5}{1.5}{}
        \node at (6, 1) {$=$};
        \node at (6.5, 1) {$0$};

        \node [left] at (0,-1) {(c)};
        \Hdots{0}{-1.5}
        \MPSTriaL{1}{-1}{}
        \MPSCirc{2}{-1}{\footnotesize$B_i^\prime$}
        \MPSTriaR{3}{-1}{}
        \AMPSTriaL{1}{-2}{}
        \AMPSCirc{2}{-2}{\footnotesize$B_i^\dagger$}
        \AMPSTriaR{3}{-2}{}
        \Hdots{4}{-1.5}
        \node at (4.5, -1.5) {$=$};
        \BEnvL{5.5}{-2}{-1}{}
        \MPSCirc{6}{-1}{\footnotesize$B_i^\prime$}
        \AMPSCirc{6}{-2}{\footnotesize$B_i^\dagger$}
        \BEnvR{6.5}{-2}{-1}{}

        \node [left] at (0,-3) {(d)};
        \Hdots{0}{-3.5}
        \MPSTriaL{1}{-3}{}
        \MPSCirc{2}{-3}{\footnotesize$B_j$}
        \AMPSTriaL{1}{-4}{}
        \AMPSTriaL{2}{-4}{}
        \Hdots{3}{-3.5}
        \MPSTriaR{4}{-3}{}
        \MPSTriaR{5}{-3}{}
        \AMPSCirc{4}{-4}{\footnotesize$B_i^\dagger$}
        \AMPSTriaR{5}{-4}{}
        \Hdots{6}{-3.5}
        \draw [line width = 1] (2.5, -3) to (3.5, -3);
        \draw [line width = 1] (2.5, -4) to (3.5, -4);        

        \node at (6.5,-3.5) {$=$};
        \node at (7,-3.5) {$0$};
\end{tikzpicture}
\caption{(a, b) show the left and right orthogonal conditions, respectively. 
(c) illustrates that the pullback inner product between two local tensors on 
the same site is just the standard Euclidean inner product of the local tensors. (d) shows that the inner product between two different sites 
vanishes because of the orthogonal gauge fixing condition.}
\label{Fig_PBIP}
\end{figure}

In the MPS representation of tangent vectors, there also exists gauge 
redundancy in tangent space that needs to be fixed. Here we adopt the 
orthogonal condition
\begin{equation}
    \left\{
    \begin{aligned}
        &\sum_{s_i}{A_i^{s_i}}^\dagger B_i^{s_i} = 0,\quad &(\text{left})\\
        &\sum_{s_i}{B_i^{s_i}A_i^{s_i}}^\dagger  = 0,\quad &(\text{right})\\
    \end{aligned}
    \right.
\end{equation} 
to fix the gauge and facilitate the optimization process later.
This is shown in Fig.~\ref{Fig_PBIP}(a-b). With this we find 
\begin{equation}
    \braket{\Phi_A^i(B_i)}{\Phi_A^j(B_j^\prime)} 
    = \delta_{ij}\sum_{s_i}\Tr{{B_i^\dagger}^{s_i}{B_i^\prime}^{s_i}} 
    = \delta_{ij} \langle B_i, B_i^\prime \rangle,
\label{EqSM:Pullback}
\end{equation}
i.e., the pullback inner product in parameter space $T_{A}\mathcal{A} 
\simeq \bigoplus_{i=1}^{N}\mathbb{C}^{D_{i-1}D_{i}\tilde{d}_i}$ is
nothing but the standard Euclidean one, as 
illustrated in Fig.~\ref{Fig_PBIP}(c-d). At the same site it is just a full 
contraction of two local tensors that results in a scaler; on the contrary, 
at different sites the inner product simply vanishes.

\subsection{Optimization of tangent vector within the tangent space \label{Sec_OptimalB}}
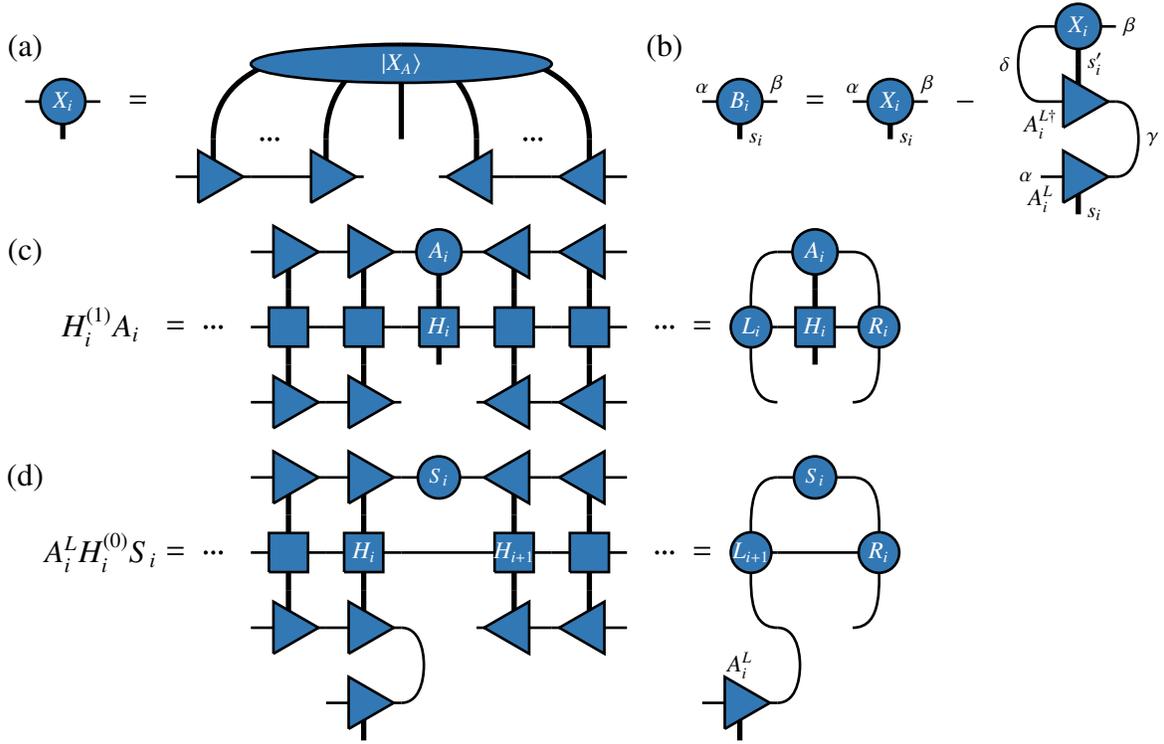
\begin{figure}[htbp]
    \centering
    \begin{tikzpicture}[scale = 1]
        \node [below] at (-1.5, 4) {\large(a)};

        \MPSCirc{-1}{3}{$X_i$}
        \node at (0, 3) {\large$=$};
        
        \draw [line width = 2] (1, 2.5) to [out = 90, in = 180] (3.5, 3.75);
        \draw [line width = 2] (2.5, 2.5) to [out = 90, in = 195] (3.5, 3.75);
        \draw [line width = 2] (3.5, 2.5) to [out = 90, in = -90] (3.5, 3.75);
        \draw [line width = 2] (4.5, 2.5) to [out = 90, in = -15] (3.5, 3.75);
        \draw [line width = 2] (6, 2.5) to [out = 90, in = 0] (3.5, 3.75);
        \draw [line width = 1, fill = DarkBlue] (3.5, 3.5) ellipse [x radius = 2, y radius = 0.25];
        \node[white] at (3.5, 3.5) {$\ket{X_A}$};
        
        \AMPSTriaL{1}{2}{}
        \AMPSTriaL{2.5}{2}{}
        
        \AMPSTriaR{4.5}{2}{}
        \AMPSTriaR{6}{2}{}
        \draw [line width = 1] (1.5, 2) to (2, 2);
        \draw [line width = 1] (5, 2) to (5.5, 2);
        \Hdots{1.75}{2.5}
        \Hdots{5.25}{2.5}   
        
        \node [below] at (7, 4) {\large(b)};
        \MPSCirc{8}{3}{$B_i$}
        \node [above] at (7.5, 3) {\footnotesize$\alpha$};
        \node [above] at (8.5, 3) {\footnotesize$\beta$};
        \node [right] at (8, 2.5) {\footnotesize$s_i$};
        \node at (9, 3) {\large$=$};

        \MPSCirc{10}{3}{$X_i$} 
        \node [above] at (9.5, 3) {\footnotesize$\alpha$};
        \node [above] at (10.5, 3) {\footnotesize$\beta$};
        \node [right] at (10, 2.5) {\footnotesize$s_i$};
        \node at (11, 3) {\large$-$};

        \MPSCirc{12.5}{4}{$X_i$}
        \AMPSTriaL{12.5}{3}{}
        \node at (12, 2.7) {$A_i^{L\dagger}$};
        \MPSTriaL{12.5}{2}{}
        \node at (12, 1.7) {$A_i^L$};
        \BEnvL{12}{3}{4}{\footnotesize$\delta$}
        \BEnvR{13}{2}{3}{\footnotesize$\gamma$}
        \node [right] at (13, 4) {\footnotesize$\beta$};
        \node [right] at (12.5, 3.5) {\footnotesize$s_i^\prime$};
        \node [right] at (12.5, 1.5) {\footnotesize$s_i$};
        \node [left] at (12, 2) {\footnotesize$\alpha$};

        \node at (-1.5, 1) {\large(c)};
        \node at (-0.5, 0) {\large $H_i^{(1)}A_i$};
        \node at (0.5, 0) {\large$=$};
        \Hdots{1}{0}

        \MPSTriaL{2}{1}{}
        \MPOSqua{2}{0}{}
        \AMPSTriaL{2}{-1}{}
        \MPSTriaL{3}{1}{}
        \MPOSqua{3}{0}{}
        \AMPSTriaL{3}{-1}{}
        \MPSCirc{4}{1}{$A_i$}
        \MPOSqua{4}{0}{$H_i$}
        \MPSTriaR{5}{1}{}
        \MPOSqua{5}{0}{}
        \AMPSTriaR{5}{-1}{}
        \MPSTriaR{6}{1}{}
        \MPOSqua{6}{0}{}
        \AMPSTriaR{6}{-1}{}
        \Hdots{7}{0}
        \node at (7.5, 0) {\large$=$};

        \TEnvLT{8.5}{-1}{0}{1}{$L_i$}
        \MPSCirc{9}{1}{$A_i$}
        \MPOSqua{9}{0}{$H_i$}
        \TEnvRT{9.5}{-1}{0}{1}{$R_i$}

        \node at (-1.5, -2) {\large(d)};
        \node at (-0.5, -3) {\large $A_i^LH_i^{(0)}S_i$};
        \node at (0.5, -3) {\large$=$};
        \Hdots{1}{-3}

        \draw [line width = 1] (3.5, -2) -- (4.5, -2);
        \draw [line width = 1] (3.5, -3) -- (4.5, -3);
        \MPSTriaL{2}{-2}{}
        \MPOSqua{2}{-3}{}
        \AMPSTriaL{2}{-4}{}
        \MPSTriaL{3}{-2}{}
        \MPOSqua{3}{-3}{$H_i$}
        \AMPSTriaL{3}{-4}{}
        \MPSTriaL{3}{-5}{}
        \BEnvR{3.5}{-5}{-4}{}
        \Circ{4}{-2}{$S_i$}
        \MPSTriaR{5}{-2}{}
        \MPOSqua{5}{-3}{$H_{i+1}$}
        \AMPSTriaR{5}{-4}{}
        \MPSTriaR{6}{-2}{}
        \MPOSqua{6}{-3}{}
        \AMPSTriaR{6}{-4}{}
        \Hdots{7}{-3}
        \node at (7.5, -3) {\large$=$};

        \draw [line width = 1] (8.5, -2) -- (9.5, -2);
        \draw [line width = 1] (8.5, -3) -- (9.5, -3);
        \TEnvLT{8.5}{-4}{-3}{-2}{$L_{i+1}$}
        \TEnvRT{9.5}{-4}{-3}{-2}{$R_i$}      
        \Circ{9}{-2}{$S_i$}  
        \MPSTriaL{8}{-5}{}
        \node at (8, -4.5) {$A_i^L$};
        \BEnvR{8.5}{-5}{-4}{}

\end{tikzpicture}
\caption{(a) shows the definition of local component $X_i$ of a tangent vector 
$\ket{X_A}$. (b) shows the tensor network representation of optimal tensor 
parameter $B_i$. (c) shows the local component $X_i = H_i^{(1)} A_i$ for 
the tangent vector $\ket{X_A} = H\ket{\Psi(A)}$, where $L_i$ and $R_i$ are
the rank-3 left and right environment tensors of the $i$-th site, respectively. 
Substitute $X_i$ in the second term of (b), and decompose $A_i = A_i^L S_i$, 
we arrive at the tensor structure in (d), which represents $A_i^LH_i^{(0)}S_i$.
}
\label{Fig_optimalB}
\end{figure}

Let $\ket{X_A} \in T_{\ket{\Psi(A)}}\mathcal{H}$ be any tangent vector 
in the Hilbert space, its local component projected onto the tangent 
space and on the $i$-th site can be defined as
\begin{equation}
X_i = \frac{\partial}{\partial B_{i}^\dagger}\braket{\Phi_A^i(B_i)}{X_A}
\label{EqSM:Xi}
\end{equation}
shown in Fig.~\ref{Fig_optimalB}(a), where $\braket{\Phi_A^i(B_i)}{X_A} 
= \langle B_i, X_i \rangle$ under the mixed gauge condition. The 
optimization problem amounts to minimize the distance
\begin{equation}
\left\|\Phi_{A}(B) - X_A\right\|^2 = \sum_i\left(\langle B_i, B_i\rangle 
- \langle B_i, X_i\rangle - \langle X_i, B_i\rangle \right) 
+ \left\|X_A\right\|^2.
\end{equation}
Without loss of generality, we choose the left-orthogonal condition
in the tangent space [c.f., Fig.~\ref{Fig_PBIP}(a)] and optimize the 
parameter $B_i$ to minimize the Lagrangian
\begin{equation}
\mathcal{L} = \langle B_i, B_i\rangle - \langle B_i, X_i\rangle - \langle X_i, B_i\rangle 
+ \sum_{\alpha,\beta}\lambda_{\alpha\beta} \sum_{s_i,\gamma}
\overline{(B_i)_{\gamma\alpha}^{s_i}}(A_i^L)_{\gamma\beta}^{s_i},
\end{equation}
where $\lambda_{\alpha\beta}$'s denote the $D_i^2$ multiplers imposing the 
left-orthogonal condition. The solution of this conditional extremum problem is
\begin{equation}
    \left(B_i\right)_{\alpha\beta}^{s_i} = \left(X_i\right)_{\alpha\beta}^{s_i} - 
    \sum_{s_i^\prime,\gamma,\delta}\overline{\left(A_i^L\right)_{\delta\gamma}^{s_i^\prime}}
    \left(X_i\right)_{\delta\beta}^{s_i^\prime}\left(A_i^L\right)_{\alpha\gamma}^{s_i},
    \label{Eq_OptimalB}
\end{equation}
as illustrated in Fig.~\ref{Fig_optimalB}(b). Note that left-orthogonal 
condition can only be imposed to the left $N-1$ sites, and on the last 
site $N$ Eq.~(\ref{Eq_OptimalB}) reduces to 
$\left(B_N\right)_{\alpha\beta}^{s_N} = \left(X_N\right)_{\alpha\beta}^{s_N}$.

In the present problem of interest, the tangent vector is generated 
by $\ket{X_A} = -H\ket{\Psi(A)}$, and the first term $X_i$ in 
Eq.~(\ref{Eq_OptimalB}) is a result of 1-site effective Hamiltonian 
$H_i^{(1)}$ acting on the local tensor $A_i$, i.e., $-H_i^{(1)}A_i$ 
[c.f., Fig.~\ref{Fig_optimalB}(c)]. Regarding the second term in 
Eq.~(\ref{Eq_OptimalB}), we firstly take a tensor decomposition 
$A_i = A_i^L S_i$ via, e.g., the QR decomposition, to move the 
MPS from a site canonical form to a bond canonical one. The second 
term of Eq.~(\ref{Eq_OptimalB}) thus becomes $A_i^L H_i^{(0)} S_i$, 
where $H_i^{(0)}$ denotes the bond effective Hamiltonian [c.f., Fig.
\ref{Fig_optimalB}(d)]. Therefore, pulled back via the tangent map 
(\ref{Eq_tangentMap}), the flow equation of local tensor $A_i$ 
reads
\begin{equation}
\frac{dA_i}{d\beta} = - H_{i}^{(1)}A_i + A_i^LH_i^{(0)}S_i,
\label{Eq_local}
\end{equation}
which is right the Eq.~(2) of the main text, whose integration 
generates the required solution $A_i(\beta)$.

\subsection{Splitting method and Lie-Trotter errors\label{Sec_LieTrotter}}
Substituting Eq.~(\ref{Eq_local}) into the tangent map Eq.~(\ref{Eq_tangentMap}), 
the optimal tangent vector reads
\begin{equation}
    \frac{d}{d\beta}\ket{\Psi(A)} = \sum_{i = 1}^N  \ket{\Phi_A^i(-H_{i}^{(1)}A_i)} + 
    \sum_{i = 1}^{N-1} \ket{\Phi_A^i(A_{i}^LH_{i}^{(0)}S_i)},
    \label{Eq_global}
\end{equation}
reemphasize that the left(right)-orthogonal condition is imposed on each site but the last(first) one. Each term of this 
summation constitutes a linear equation as $H_i^{(1)}$ ($H_i^{(0)}$) independent on $A_i$ ($S_i$), 
while the combination as a whole is nonlinear. A standard technique to integrate it 
is the splitting method~\cite{GeometricIntegral}, i.e., we split the tangent vector field corresponding to Eq.~(\ref{Eq_global}) into the $2N-1$ tangent vector fields, and
arrive at the following linear equations
\begin{equation}
    \frac{d}{d\beta}\ket{\Psi(A)} = \ket{\Phi_A^i(-H_{i}^{(1)}A_i)}
    \quad\text{and}\quad 
    \frac{d}{d\beta}\ket{\Psi(A)} = \ket{\Phi_A^i(A_{i}^LH_{i}^{(0)}S_i)}
\end{equation} 
that can be successively integrated one by one. With tensor-network parameterization, 
the linear equations of local tensors read 
\begin{equation}
\label{EqS:dAdb}
    \frac{dA_i}{d\beta} = - H_{i}^{(1)}A_i
    \quad\text{and}\quad 
    \frac{dS_i}{d\beta} = H_{i}^{(0)}S_i,
\end{equation} 
which can be exactly solved as $A_i(\beta + \tau) = e^{-\tau H_i^{(1)}}A_i(\beta)$ 
and $S_i(\beta + \tau) = e^{\tau H_i^{(0)}}S_i(\beta)$. Generically, such a splitting 
leads to Lie-Trotter errors that scale as $O(\tau)$, which, however, can be reduced 
to $O(\tau^2)$ when we choose a symmetric integrator, i.e., composition of 
left-to-right and right-to-left sweeps that are adjoint with each other
\cite{TDVP2016,GeometricIntegral,JhengWeionetdvp}.

Moreover, when restricted to the affine Hilbert space 
$T_{\ket{\Psi(A)}} \mathcal{M}$, instead of the MPS manifold $\mathcal{M}$, 
the flow equation as a whole is guarenteed to be linear. After the same splitting 
procedure, and given the orthogonal conditions, it resorts to solving the system 
consisted of $2N-1$ effective Hamiltonians that are mutually commutative, and 
the Lie-Trotter splitting becomes exact in this case. Back to the standard 1-TDVP 
on manifold $\mathcal{M}$, we argue that the Lie-Trotter error decreases as the 
bond dimension $D$ is increased, as the manifold $\mathcal{M}$ locally resembles 
the tangent space $T_{\ket{\Psi(A)}}\mathcal{M}$ in the large $D$ limit. Therefore, 
the Lie-Trotter error, as well as the projection error, is well controlled by the parameter 
$D$, and this constitutes a very promising feature of tangent-space based approach
of tensor networks.

\subsection{Procedure of 1-TDVP \label{Sec_procedure}}
We describe the comprehensive procedure of 1-TDVP integrator, 
adapted for the imaginary-time evolution of super MPS $\kett{\rho}$, 
in the practical level as follows. 
\begin{enumerate}
\item Starting from the leftmost site, i.e. $i=1$, which is also the canonical 
center of the super MPS, with the left-orothogonal condition of tangent vector 
[Fig.~\ref{Fig_PBIP}(a)] is considered.
\item Integrate the first term of Eq.~(\ref{Eq_local}) and obtain $A(\beta_0 + \tau) 
=e^{-H_i^{(1)}\tau}A_i(\beta_0)$, where $\tau$ is the step length.
\item Take a decomposition $A_i = A_i^L S_i$ via the QR or singular value 
decomposition (SVD), and gauge the super MPS $\kett{\rho}$ to a bond 
canonical form centered at $i$-th bond.
\item Integrate the second term of Eq.~(\ref{Eq_local}) and update the 
bond tensor $S_i(\beta_0+\tau) = e^{H_i^{(0)}\tau}S_i(\beta_0)$.
\item Contract the updated bond tensor $S_i(\beta_0+\tau)$ with the 
local tensor $A_{i+1}^R$ to the right of bond $i$, thus moving the canonical 
center to site $i+1$.
\item Repeat steps $2$-$5$ until the canonical center arrives at the rightmost 
site $N$, where no bond update is needed to the right of this site.
\item Sweep backwards from right to left, following the same line as the forward 
sweep and with the right-orthogonal condition [Fig.~\ref{Fig_PBIP}(b)] imposed 
instead. 
\end{enumerate}

\subsection{2-TDVP in the initial stage \label{Sec_2TDVP}}
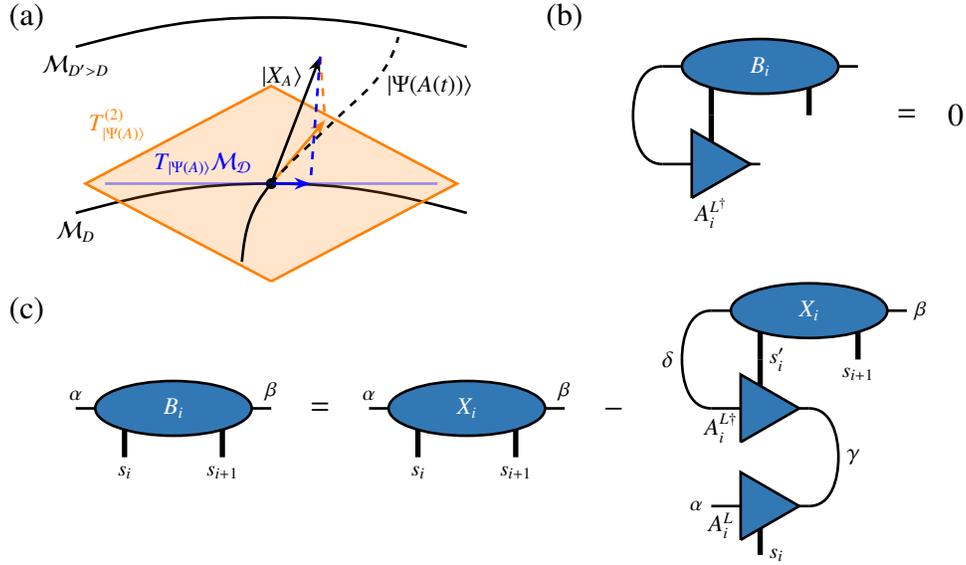
\begin{figure}[htbp]
\begin{tikzpicture}[scale = 1.3]

    \node at (0, 2) {\large (a)};
    \draw [line width = 1] (0.5, 1.7) to [out = 15, in = 180] (2.5, 2) to [out = 0, in = 165] (4.5, 1.7);
    \node [below] at (0.5, 1.7) { $\mathcal{M}_{D^\prime > D}$};
    \draw [line width = 1] (0.5, 0) to [out = 15, in = 180] (2.5, 0.3) to [out = 0, in = 165] (4.5, 0);
    \node [below] at (0.5, 0) {$\mathcal{M}_{D}$};
    \draw [line width = 1, orange, fill = orange, fill opacity = 0.2] (2.5, -0.7) -- (4.4, 0.3) -- (2.5, 1.3) -- (0.6, 0.3) -- (2.5, -0.7);
    \draw [line width = 1, -{Stealth[length=6pt]}, orange] (2.5, 0.3) -- (3.05, 0.95);
    \draw [line width = 1, orange, dashed] (3, 1.6) -- (3.05, 0.95);
    \node [left] at (1.3, 0.9) {\color{orange}{$T_{\ket{\Psi(A)}}^{(2)} $}};
    \draw [line width = 1, -{Stealth[length=6pt]}] (2.5, 0.3) -- (3, 1.6);
    \node [left] at (2.9, 1.4) {$\ket{X_A}$};
    \draw [fill] (2.5, 0.3) circle [radius = 0.05];
    \draw [line width = 1, blue, opacity = 0.4] (0.8, 0.3) -- (4.2, 0.3);
    \draw [line width = 1, -{Stealth[length=6pt]}, blue] (2.5, 0.3) -- (2.9, 0.3);
    \draw [line width = 1, blue, dashed] (3, 1.6) -- (2.9, 0.3);
    \node [above] at (1.8, 0.3) {\color{blue}{$T_{\ket{\Psi(A)}}\mathcal{M_{D}}$}};
    \draw [line width = 1] (2.2, -0.5) to [out = 85, in = 229.4] (2.5, 0.3);
    \draw [line width = 1, dashed] (2.5, 0.3) to [out = 49.4, in = -135] (3.1, 0.9) to [out = 45, in = -105] (3.8, 1.8);
    \node [right] at (3.6, 1.3) {$\ket{\Psi(A(t))}$};

    \node at (5.5, 2) {\large (b)};
    \MPSEllipse{7.5}{1.5}{$B_i$}
    \AMPSTriaL{7}{0.5}{}
    \node at (7, 0) {$A_i^{L^\dagger}$};
    \BEnvL{6.5}{0.5}{1.5}{}
    \node at (9,1) {\large$=$};
    \node at (9.5, 1) {\large$0$};

    \node at (0, -1) {\large (c)};
    \MPSEllipse{1.5}{-2}{$B_i$}
    \node [above] at (0.5, -2) {\footnotesize $\alpha$};
    \node [above] at (2.5, -2) {\footnotesize  $\beta$};
    \node [below] at (1, -2.5) {\footnotesize  $s_i$};
    \node [below] at (2, -2.5) {\footnotesize  $s_{i+1}$};
    \node at (3, -2) {\large$=$};

    \MPSEllipse{4.5}{-2}{$X_i$} 
    \node [above] at (3.5, -2) {\footnotesize $\alpha$};
    \node [above] at (5.5, -2) {\footnotesize  $\beta$};
    \node [below] at (4, -2.5) {\footnotesize  $s_i$};
    \node [below] at (5, -2.5) {\footnotesize  $s_{i+1}$};
    \node at (6, -2) {\large$-$};

    \MPSEllipse{8}{-1}{$X_i$}
    \AMPSTriaL{7.5}{-2}{}
    \node at (7.1, -2.2) {$A_i^{L\dagger}$};
    \MPSTriaL{7.5}{-3}{}
    \node at (7.1, -3.2) {$A_i^L$};
    \BEnvL{7}{-2}{-1}{$\delta$}
    \BEnvR{8}{-3}{-2}{$\gamma$}
    \node [right] at (9, -1) {\footnotesize $\beta$};
    \node [right] at (7.5, -1.5) {\footnotesize $s_i^\prime$};
    \node [below] at (8.5, -1.5) {\footnotesize $s_{i+1}$};
    \node [right] at (7.5, -3.5) {\footnotesize $s_i$};
    \node [left] at (7, -3) {\footnotesize $\alpha$};

\end{tikzpicture}
\caption{(a) illustrates the comparison between 2-TDVP and 1-TDVP. Note 
that we plot the manifold $\mathcal{M}_D$ of MPS with bond dimension 
$D$ as a 1D manifold for the sake of convenience, and its tangent space 
$T_{\ket{\Psi(A)}}\mathcal{M_{D}}$ is depicted by a blue line. The space 
of 2-site variations $T_{\ket{\Psi(A)}}^{(2)}$ is a linear subspace of 
$T_{\ket{\Psi(A)}}\mathcal{H}$ which is strictly larger than $T_{\ket{\Psi(A)}} 
\mathcal{M_{D}}$, represented by the orange plane. The black arrow 
denotes the exact tangent vector $\ket{X_A} \in T_{\ket{\Psi(A)}}\mathcal{H}$, 
and the blue and orange arrows are the projection to $T_{\ket{\Psi(A)}}
\mathcal{M_{D}}$ or $T_{\ket{\Psi(A)}}^{(2)}$, respectively. In 1-TDVP, 
the quantum-state/density-operator flow is induced by the blue tangent 
vector and hence the state cannot leave the manifold $\mathcal{M}_D$. 
However, the orange tangent vector has components normal to 
$T_{\ket{\Psi(A)}} \mathcal{M_{D}}$, hence the flow in 2-TDVP 
can leave the manifold $\mathcal{M}_D$ and explore a manifold of larger 
bond dimensions, as illustrated by the dashed black curve. (b) shows the 
left-orthogonal condition of local tensor $B_i$ in the 2-TDVP, and (c) shows 
the tensor network representation of the determined optimal two-site tensor $B_i$.
}
\label{Fig_2tdvp}  
\end{figure}

As the imaginary-time evolution of MPO $\rho$ (or equivalently the super 
MPS $\kett{\rho}$) starts from a high-temperature initial state $\rho(\tau_0) 
= 1 -\tau_0 H$ whose bond dimension is relatively small, 
as $D = D_H$ (bond dimension of the Hamiltonian MPO). It is therefore 
necessary to devise an algorithm that can increase the bond dimension 
adaptively. The 1-TDVP algorithm projects the exact tangent vector 
$\ket{X_A}$ into the (MPS) tangent space $T_{\ket{\Psi(A)}}\mathcal{M}$, 
which is a subspace of $T_{\ket{\Psi(A)}}\mathcal{H}$. Inspired by this, 
we choose a larger subspace, which contains the components normal to 
the MPS tangent space, and hence can go beyond the manifold $\mathcal{M}$. 
Such an algorithm dubbed as 2-TDVP using two-site update is illustrated 
in Fig.~\ref{Fig_2tdvp}(a). The idea is straightforward, we ``glue'' two 
physical indices of a pair of nearest sites, and the bond dimension is 
increased adaptively when we separate the two sites again after the evolution. 

To be specific, in 2-TDVP the projected tangent vector becomes  
\begin{equation}
\widetilde{\Phi}_A(B) = \sum_{i = 1}^{N-1} \widetilde{\Phi}_A^i(B_i)
= \sum_{i = 1}^{N-1}\sum_{\{s_j\}}\Tr{{A^L}_1^{s_1}\cdots {{A^L}_{i-1}^{s_{i-1}} 
B_{i}^{s_i,s_{i+1}} {A^R}_{i+2}^{s_{i+2}}\cdots {A^R}_N^{s_N}}}\ket{s_1\cdots s_N},
\end{equation}
where $B = (B_1, \cdots, B_{N-1})$ and each $B_i$ is now a rank-4 local 
tensor related to a pair of nearest sites $i$ and $i+1$. The linear subspace 
corresponding to 2-TDVP is exactly the space of all the two-site variations, 
denoted by $T_{\ket{\Psi(A)}}^{(2)}$. It can be checked that any one-site
variation that spans the tangent space $T_{\ket{\Psi(A)}}\mathcal{M}$ can 
be regarded as a special two-site variation. In this sense, $T_{\ket{\Psi(A)}}^{(2)}$ 
is indeed a larger subspace, i.e., we have the following rigorous relation
\begin{equation}
T_{\ket{\Psi(A)}}\mathcal{M}\subset T_{\ket{\Psi(A)}}^{(2)} \subset T_{\ket{\Psi(A)}}\mathcal{H}.
\end{equation}

Following the same line as in the derivation of 1-TDVP, we introduce the 
left-orthogonal condition of tangent vectors as
\begin{equation}
\sum_{s_i}{A_i^{s_i}}^\dagger B_i^{s_i,s_{i+1}} = 0,
\end{equation}
which is illustrated in Fig.~\ref{Fig_2tdvp}(b). 
The corresponding Lagrangian now reads
\begin{equation}
\mathcal{L} = \langle B_i, B_i\rangle - \langle B_i, X_i\rangle 
- \langle X_i, B_i\rangle + \sum_{\alpha,\beta,s_{i+1}} 
\lambda_{\alpha\beta}^{s_{i+1}} \sum_{s_i,\gamma}
\overline{(B_i)_{\gamma\alpha}^{s_is_{i+1}}}(A_i^L)_{\gamma\beta}^{s_i},
\end{equation}
where $X_i$ is a two-site tensor with the same size of $B_i^{s_i,s_{i+1}}$. 
Solve this extremum value problem and we arrive at
\begin{equation}
\left(B_i\right)_{\alpha\beta}^{s_is_{i+1}} = \left(X_i\right)_{\alpha\beta}^{s_is_{i+1}} - 
\sum_{s_i^\prime,\gamma,\delta}\overline{\left(A_i^L\right)_{\delta\gamma}^{s_i^\prime}}
\left(X_i\right)_{\delta\beta}^{s_i^\prime s_{i+1}}\left(A_i^L\right)_{\alpha\gamma}^{s_i},
\end{equation}
see Fig.~\ref{Fig_2tdvp}(c). If the tangent vector is induced by imaginary-time 
evolving equation in particular, then the flow equation of local tensors reads
\begin{equation}
    \frac{dC_i}{d\beta} = - H_{i}^{(2)}C_i + A_i^LH_{i+1}^{(1)}A_{i+1},
    \label{EqSM:2TDVP}
\end{equation}
where $C_i = A_i^L A_{i+1}$ is the local two-site tensor and $H_{i}^{(2)}$ denotes 
the two-site effective Hamiltonian acting on site $i$ and $i+1$. Overall, instead of 
$N$ 1-site update plus $(N-1)$ bond update operations as in 1-TDVP, the 
2-TDVP now involves $(N-1)$ 2-site updates and $(N-2)$ 1-site updates. 

\subsection{Lanczos-based exponential method \label{Sec_Lanczos}}
Take the 1-TDVP as an example, after the splitting in Sec.~\ref{Sec_LieTrotter}, 
we need to integrate the flow equation Eq.~(\ref{EqS:dAdb}) and compute 
$A_i(\beta) = e^{-H_{i}^{(1)}(\beta - \beta_0)}A_i(\beta_0)$ and 
$S_i(\beta) = e^{H_{i}^{(0)}(\beta - \beta_0)}S_i(\beta_0)$.
Note that $H_{i}^{(1)}$ is a rank-6 tensor acting on the rank-3 local tensor $A_i$ 
and can be regarded as a $D_{i-1}D_{i} \tilde{d}_i\times D_{i-1}D_{i} \tilde{d}_i$ 
matrix. The computational complexity of brute-force exponential method is 
very high and scales as $O(D^6)$, which can be greatly reduced to $O(D^3)$ 
via the Lanczos exponential technique as we actually only need to compute 
$e^{-H_{i}^{(1)}(\beta - \beta_0)} A_i(\beta_0)$ instead of $e^{-H_{i}^{(1)}(\beta 
- \beta_0)}$ explicitly. 

The details of Lanczos-based exponential method are as follows. We take $b$ 
as the initial state in a Hilbert space and $H$ is an Hermitian operator acting on 
it. The key point of Lanczos exponential is to project the operator to the Krylov 
space $\mathcal{K} \equiv {\rm span}\{b, Hb, \cdots, H^{K-1}b\}$, where $K$ 
the Krylov space dimension. Through the standard Lanczos procedure, we find a group of orthogonal basis $\{q_i\}_{i = 1}^{K}$ and the representation 
$H \approx QTQ^\dagger$ where $Q = [q_1, \cdots, q_K]$ and $T$ is a tridiagonal 
matrix. Given the Krylov space constructed, the exponential can be calculated 
as $e^{H}b \approx Qe^TQ^\dagger b = \|b\|\sum_{i = 1}^K c_i q_i$ where the 
coefficients $c_i$'s constitute the first column of $e^T$. Note $b$ is the initial vector 
of the Lanczos procedure and $q_1 = b/\|b\|$.

\begin{figure}[htbp]
\begin{tikzpicture}[scale = 1.1]
        \TEnvLT{0}{-1}{0}{1}{$L_i$}
        
        \draw [line width = 1, ->] (0.5,0) -- (1.5, 0); 
        \node [above] at (1, 0) {$D^3D_H$};

        \TEnvLT{2.5}{-1}{0}{1}{$L_i$}
        \MPSCirc{3}{1}{$A_i$}

        \draw [line width = 1, ->] (4,0) -- (5, 0); 
        \node [above] at (4.5, 0) {$D^2D_H^2$};

        \TEnvLT{6}{-1}{0}{1}{$L_i$}
        \MPSCirc{6.5}{1}{$A_i$}
        \MPOSqua{6.5}{0}{$H_i$}

        \draw [line width = 1, ->] (7.5,0) -- (8.5, 0); 
        \node [above] at (8, 0) {$D^3D_H$};
        
        \TEnvLT{9.5}{-1}{0}{1}{$L_i$}
        \MPSCirc{10}{1}{$A_i$}
        \MPOSqua{10}{0}{$H_i$}
        \TEnvRT{10.5}{-1}{0}{1}{$R_i$}
\end{tikzpicture}
\caption{The tensor contraction of $H_{i}^{(1)}A_i$ with $O(D^3)$ 
complexity in 1-site update.}
\label{Fig_Lanczos}
\end{figure}
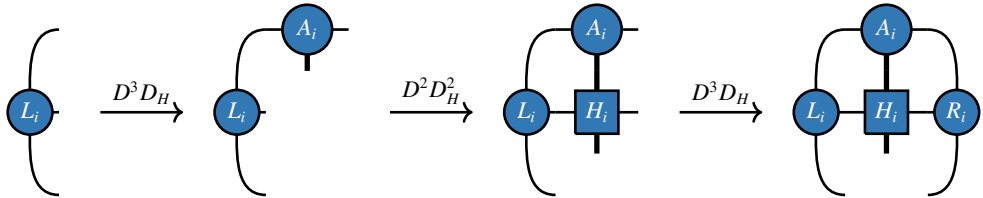

In the Lanczos procedure, we need to apply $H: v\mapsto Hv$ iteratively, whose 
computational complexity is $O(D^3)$. In Fig.~\ref{Fig_Lanczos} we illustrate the 
details of the tensor contractions involved in the 1-site update: we firstly contract 
$L_i$ and $A_i$, then apply $H_i$ to it, and finally contract the resultant with $R_i$. 
The complexity is thus estimated as
\begin{equation}
    O(D^3D_H) + O(D^2D_H^2) + O(D^3D_H) = O(D^3D_H).
\end{equation}
Note that $D_H$ is the bond dimensions of the Hamiltonian MPO, which is much smaller than $D$ in most 
cases involved in the present work. Besides 1-site update, the processes of bond 
and 2-site updates are similar and both of them are also with $O(D^3)$ complexity.  

\section{Determinant Quantum Monte Carlo Simulations}

\begin{figure}[htbp]
\centering
\includegraphics[width = .8\linewidth]{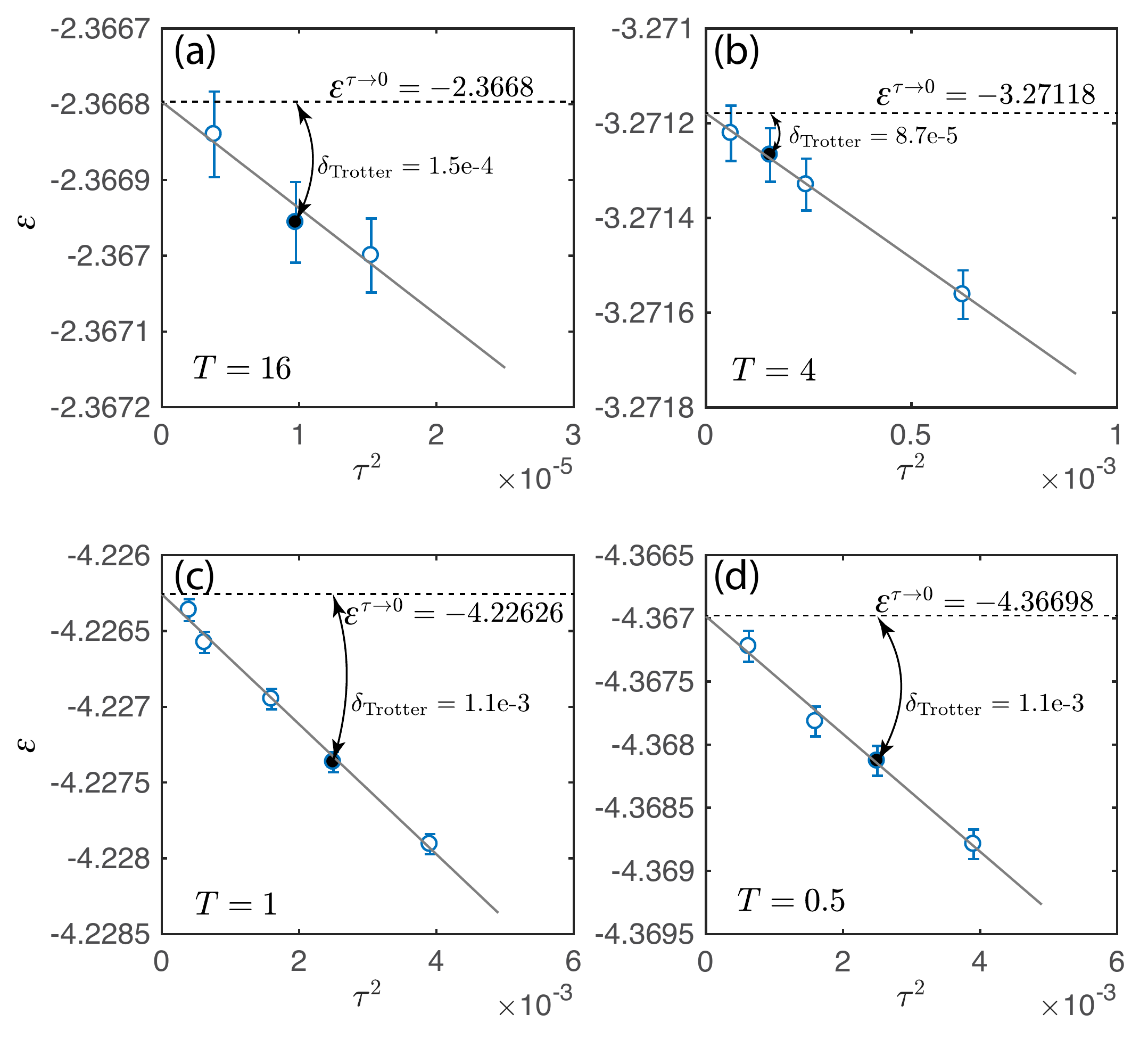}
\caption{In a CL$8\times16$ system with $U=8$, DQMC energies density 
$\varepsilon$ are shown versus the square of Trotter slices $\tau^2$ for (a) $T/t=16$
(b) $T/t=4$, (c) $T/t=1$, (d) $T/t=0.5$, from which linear extrapolations on 
$\tau^2$ are performed. 
For the data shown in the main text (solid dots), the corresponding Trotter errors are
$\delta_\mathrm{Trotter}=1.5\times10^{-4}, 8.7\times10^{-5}, 1.1\times10^{-3}, 1.1\times10^{-3}$.}
\label{Fig_S11}    
\end{figure} 

\begin{figure}[htbp]
\centering
\includegraphics[width = .5\linewidth]{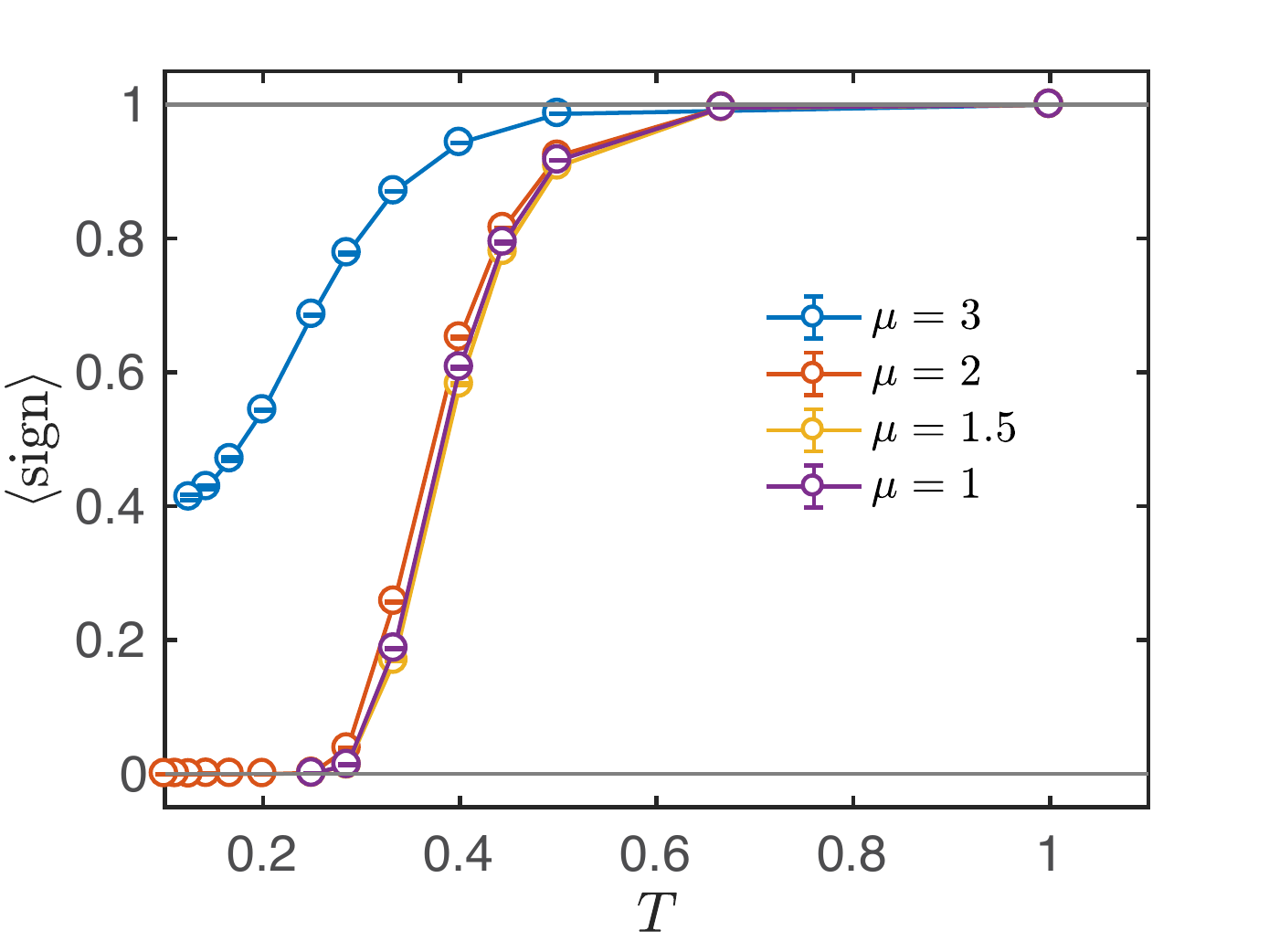}
\caption{In a CL$8\times16$ system with $U=8$, the average signs $\langle\mathrm{sign}\rangle$ 
are shown versus $T$ for different chemical potentials $\mu$, which show rapid decay to 0 
around $T=0.4$ for the cases of $\mu=1, 1.5, 2$. }
\label{Fig_S12}    
\end{figure} 

In this section, we briefly recapitulate the determinant quantum Monte Carlo (DQMC)
\cite{Blankenbecler1981,Hirsch1983,Hirsch1985,Assaad2008} method involved in the 
calculations. Below, we consider the square-lattice Hubbard model $H = H_0 + H_I$, 
where the kinetic energy $H_0$ and the on-site repulsion term $H_I$ express as
\begin{equation}
H_0 = -t\sum_{\langle i,j\rangle,\sigma}( c^\dag_{i,\sigma} c^{\,}_{j,\sigma} + \mathrm{H.c.}) 
= -t\sum_{i,j,\sigma} c^\dag_{i,\sigma} K^{\,}_{ij} c^{\,}_{j,\sigma}, \quad \quad \quad
H_I = U\sum_i (n_{i\uparrow}-\tfrac{1}{2})(n_{i\downarrow}-\tfrac{1}{2}),
\end{equation}
with the matrix elements $K_{ij}\neq0$ only if $i$ and $j$ are nearest-neighbor sites.
The thermal density operator at inverse temperature $\beta\equiv1/T$ can be approximated
via the Trotter-Suzuki decomposition \cite{Trotter1959,Suzuki1976} as, 
\begin{equation}
\rho=e^{-\beta H} = (e^{-\tau H})^M \approx (e^{-\tau H_0} e^{-\tau H_I})^M,
\end{equation} 
where small imaginary time slice $\tau=\beta/M$ is taken to ensure sufficiently small 
Trotter error $\sim \mathcal{O}(\tau^2)$ due to the fact $[H_0,H_I]\neq0$. For the 
interaction term, we make use of the discrete form of Hubbard-Stratonovich (HS) 
transformation \cite{Hirsch1983},
\begin{equation}
e^{-\tau H_I} = \prod_i e^{-\tau U (n_{i\uparrow}-\tfrac{1}{2})(n_{i\downarrow}-\tfrac{1}{2})} 
= \prod_i e^{\tau \tfrac{U}{2} (n_{i\uparrow} - n_{i\downarrow})^2 -\tau \tfrac{U}{4}}  
= \prod_i \gamma\sum_{s_i=\pm1} e^{\alpha s_i (n_{i,\uparrow} - n_{i,\downarrow})}
= \gamma^N \sum_{\{s_i=\pm1\}} e^{\sum_i \alpha s_i (n_{i,\uparrow}-n_{i,\downarrow})},
\end{equation}
with $N$ the total number of sites, $\gamma=\frac{1}{2}e^{-\tau U/4}$, 
and $\cosh{\alpha} = e^{\tau U/2}$. That is, the exponential of four-fermion 
terms are transformed to the exponential of two-fermion terms coupled 
to the (HS) ``field'' $\{s_i\}$. The many-body partition function thus writes 
\begin{align}
\mathcal{Z} &= \mathrm{Tr}\big(e^{\beta H}\big) 
\approx \mathrm{Tr}\big[(e^{-\tau H_0} e^{-\tau H_I})^M\big] 
= \gamma^{NM}  \sum_{\{s^l_i\}}  \mathrm{Tr} \left[\prod_{l=1}^M 
\left(e^{-\tau \sum_{ij,\sigma} c_{i,\sigma}^\dag K_{ij} c_{j,\sigma}^{\,}} ~ 
e^{\sum_i \alpha s^l_i (n_{i,\uparrow}-n_{i,\downarrow})}\right)\right] \\
&= \gamma^{NM}  \sum_{\{s^l_i\}}  \mathrm{Det} \left[I + \prod_{l=1}^M 
\left(e^{-\tau K} \cdot e^{\Lambda^l_\uparrow}\right)\right] \times \mathrm{Det}
\left[I + \prod_{l=1}^M \left(e^{-\tau K} \cdot e^{\Lambda^l_\downarrow}\right)\right],
\end{align}
where $I$ is an $N\times N$ identity matrix, and $\Lambda_\uparrow^l$ 
and $\Lambda_\downarrow^l$ are diagonal matrix with the $i$-th entry 
being $\alpha s_i^l$ and $-\alpha s_i^l$ respectively. 
We assign each configuration $\mathcal{C}$ of $\{s_i^l\}$ with a weight 
\begin{equation}
W_s(\mathcal{C}) =  \frac{\gamma^{NM}}{\mathcal{Z}} \mathrm{Det} 
\left[I + \prod_{l=1}^M \left(e^{-\tau K} \cdot e^{\Lambda^l_\uparrow}\right)\right] \times
\mathrm{Det} \left[I + \prod_{l=1}^M \left(e^{-\tau K} \cdot e^{\Lambda^l_\downarrow}\right)\right],
\end{equation}
and the thermodynamics of the Hubbard system is then reformulated into 
the summation over $N\times M$ Ising field $\{s_i^l\}$, with $i$ labeling 
the site and $l$ the imaginary-time slice. In Markov chain Monte Carlo, 
we consider the relative weights between configurations, and thus for each 
configuration we instead calculate 
\begin{equation}
\tilde W_s(\mathcal{C}) = \mathrm{Det} \left[I + \prod_{l=1}^M \left(e^{-\tau K} \cdot e^{\Lambda^l_\uparrow}\right)\right] \times
\mathrm{Det} \left[I + \prod_{l=1}^M \left(e^{-\tau K} \cdot e^{\Lambda^l_\downarrow}\right)\right],
\end{equation}
and use this relative weights to update Ising configurations.

To sum up, we provide the workflow of the Markov chain samplings in the DQMC below.
\begin{enumerate}
\item \textit{Initilization}---We start with an initial random configuration 
$\mathcal{C}$ of the HS field $\{s_i^l\}$ and calculate its weight $\tilde 
W_s(\mathcal{C})$.
\item \textit{Proposal of local updates}---At site $1$ and imaginary-time 
slice $1$, we flip the sign of the HS field, i.e. $(s_1^1)' = -s_1^1$. 
\item \textit{Accept/Decline}---We calculate the associated weight 
$\tilde W_s(\mathcal{C}')$, accept or decline the local update 
according to the ratio $\frac{W_s(\mathcal{C}')}{W_s(\mathcal{C}_1)}$ 
of the weights.
\item \textit{Full sweep}---We repeat step 2 and 3, for each site $i$ 
from $1$ towards $N$ and for each slice $l$ from $1$ towards $M$. 
After that, we call it a full sweep of the configuration.
\item \textit{Measurement}---For each fixed configuration of HS field $\{s_i^l\}$, 
the measurement of observable $\hat O$ can be conducted like in an
interaction-free fermion system. 
\item We repeat the full sweep process $N_\mathrm{warm}$ times 
to thermalize the systems, and after that take $N_\mathrm{measure}$ 
times ``full sweep $+$ Measurement'' procedures to collect 
$N_\mathrm{measure}$ measurements of the observables.
\end{enumerate}

In practical calculations, we set $(N_\mathrm{warm}, N_\mathrm{measure},
N_\mathrm{chain})=(1000, 5000, 300)$ for the half-filled cases and 
up to (1000, 5000, 500) for the doped cases, where $N_\mathrm{chain}$ 
is the number of Markov chains. I.e., we totally have $1,500,000$ 
measurements at half-filling cases and $2,500,000$ measurements 
at the doped cases for each observable. 
In the main text, we take $\tau=0.05$ for the $T/t\leq1$ cases, and take $\tau=\beta/20$ for 
the higer-$T$ cases, rendering the Trotter errors within the 
order ${O}(10^{-3})$. In Fig.~\ref{Fig_S11}, we explicitly show 
the energies density $u$ obtained from DQMC simulations versus the 
Trotter slice $\tau^2$ for two different temperatures $T/t=1/16, 1/4, 1, 0.5$ from panel (a) to (d). 
It can be seen that, the relative difference between the energies shown in the main text 
(indicated as the black filled dots) $\tau=0.05$ and the extrapolated values ($\tau \rightarrow 0$) are 
within the order ${O}(10^{-3})$.

\end{document}